%% file: main.tex
\pdfoutput=1
\documentclass[aps,pre,onecolumn,11pt,superscriptaddress,nofootinbib,longbibliography]{revtex4-2}
\usepackage{amsmath,amssymb}
\usepackage{graphicx}
\usepackage{booktabs}
\usepackage{bm}
\usepackage{array}
\usepackage{placeins}
\usepackage{url}
\usepackage{multirow}
\usepackage{xcolor}
\usepackage[colorlinks=true,linkcolor=blue!60!black,citecolor=blue!60!black,urlcolor=blue!60!black]{hyperref}
\hypersetup{pdftitle={Physics-constrained neural search for model discrimination and complexity reduction in seed imbibition},pdfauthor={L. Meacci, A. Farina, A. Aliyari, V. di Bari, L. Fusi, D. Fanelli}}
\graphicspath{{figures/}}
\input{numbers}

\begin{document}
\renewcommand{\baselinestretch}{1.15}\normalsize
\input{body}
\bibliography{refs}
\end{document}

%% file: numbers.tex
\newcommand{\Ntot}{64}

\newcommand{\Nweigh}{192}

\newcommand{\overLowMax}{0.18}
\newcommand{\overFiftyList}{0.92, 0.57, 0.74}

\newcommand{\aFiftyCutOne}{26.1}
\newcommand{\aFiftyCutTwo}{25.7}
\newcommand{\Esim}{33.7}
\newcommand{\residSim}{2.4}
\newcommand{\factorSim}{8}
\newcommand{\ampFifty}{2.33}
\newcommand{\spreadFourEighty}{0.75}

\newcommand{\lamVal}{0.1}
\newcommand{\nparSix}{44}
\newcommand{\nwSix}{19}

\newcommand{\EnnMin}{33.9}
\newcommand{\EnnMax}{35.1}
\newcommand{\LnetFourMin}{0.07}
\newcommand{\LnetFourMax}{0.08}
\newcommand{\LnetMidMax}{0.01}
\newcommand{\LnetFiftyMin}{0.67}
\newcommand{\LnetFiftyMax}{0.72}

\newcommand{\tGtwoWideMin}{176}
\newcommand{\tGtwoWideMax}{178}
\newcommand{\tGtwoOneNet}{73}
\newcommand{\gTwoAtTwoForty}{36}
\newcommand{\gTwoAtFourEighty}{45}
\newcommand{\gTwoAtSevenTwenty}{89}
\newcommand{\supLossFree}{12.4}

\newcommand{\LzeroMax}{3.9}
\newcommand{\spFoldsSel}{0.3}
\newcommand{\spFoldsLow}{0.7}
\newcommand{\lamNmin}{1.12}
\newcommand{\lamNmax}{1.19}
\newcommand{\lamTestMin}{0.164}
\newcommand{\lamTestMax}{0.166}
\newcommand{\lamLate}{0.1}
\newcommand{\LearlyMin}{1.2}
\newcommand{\LearlyMax}{1.3}
\newcommand{\tEarlyMax}{33}
\newcommand{\sigPool}{0.13}
\newcommand{\sRepFour}{0.07}
\newcommand{\sRepFifty}{0.20}
\newcommand{\sRepFourEighty}{0.39}

\newcommand{\wRatio}{10}
\newcommand{\wAltNmin}{1.14}
\newcommand{\wAltNmax}{1.16}
\newcommand{\wAltLmin}{0.64}
\newcommand{\wAltLmax}{0.70}

\newcommand{\nWD}{1.09}

\newcommand{\dropMin}{0.22}
\newcommand{\dropMax}{0.39}
\newcommand{\nSupMin}{1.12}
\newcommand{\nSupMax}{1.13}

\newcommand{\supBest}{0.7}
\newcommand{\supOne}{3.0}
\newcommand{\supHigh}{4.4}
\newcommand{\supDiff}{12}
\newcommand{\cvBatFO}{$+6$}
\newcommand{\cvBatFOLo}{$+1$}
\newcommand{\cvBatFOHi}{$+12$}

\newcommand{\teBatNNsix}{0.164}

\newcommand{\teBatREF}{0.162}

\newcommand{\cvRanFO}{$+7$}
\newcommand{\cvRanFOLo}{$+5$}
\newcommand{\cvRanFOHi}{$+9$}

\newcommand{\teRanNNsix}{0.142}

\newcommand{\cvTimFO}{$+6$}
\newcommand{\cvTimFOLo}{$0$}
\newcommand{\cvTimFOHi}{$+13$}

\newcommand{\teTimNNsix}{0.164}

\newcommand{\widthSpread}{5}
\newcommand{\relSig}{8}
\newcommand{\tauFour}{348}
\newcommand{\tauFifteen}{199}
\newcommand{\tauTwentyFive}{124}
\newcommand{\tauFifty}{43}
\newcommand{\nOneMinusOne}{0.14}
\newcommand{\riseIndep}{1.7}
\newcommand{\riseRel}{3}
\newcommand{\riseMean}{4}
\newcommand{\riseSD}{6}
\newcommand{\mSevenTwentyDiff}{0.03}
\newcommand{\trainConv}{0.5}
\newcommand{\NrandTest}{38}
\newcommand{\trainEnd}{0.128}
\newcommand{\cvLallMax}{2}
\newcommand{\lcMinIt}{500}

\newcommand{\cvNNCFmax}{5}
\newcommand{\cvNNCFrel}{3}

\newcommand{\NranSplits}{12}
\newcommand{\NtimeFolds}{16}

\newcommand{\kappaVal}{8.09}
\newcommand{\Efin}{33.9}
\newcommand{\nOne}{1.14}
\newcommand{\nOneLong}{1.143}
\newcommand{\LFifty}{0.67}
\newcommand{\UFifty}{2.26}

\newcommand{\thetaOn}{4.07}
\newcommand{\thetaTwo}{8.94}

\newcommand{\tOn}{180}
\newcommand{\tauTwo}{384}

\newcommand{\npar}{6}
\newcommand{\nparWord}{six}
\newcommand{\rmseFin}{0.084}
\newcommand{\rmseNet}{0.074}

\newcommand{\nOneLoNF}{1.06}
\newcommand{\nOneHiNF}{1.24}

\newcommand{\tOnLo}{145}
\newcommand{\tOnHi}{298}

\newcommand{\tauTwoLo}{195}
\newcommand{\tauTwoHi}{637}

\newcommand{\Bboot}{400}

\newcommand{\LallHiFour}{0.24}
\newcommand{\LallHiFifteen}{0.08}
\newcommand{\LallHiTwentyFive}{0.15}

\newcommand{\devRefRed}{1}
\newcommand{\earlyNone}{1.12}
\newcommand{\earlyTon}{27--40}
\newcommand{\earlyL}{0.74--0.76}
\newcommand{\convNone}{1.12}
\newcommand{\convTon}{176--177}
\newcommand{\convL}{0.71--0.72}
\newcommand{\TabData}{%
4 & 5.05 & 6.22 & 8.99 & 1440 & 8.87 & 0.06 & 2.65 \\
15 & 5.10 & 6.28 & 8.81 & 720 & 8.78 & 0.05 & 2.49 \\
25 & 5.07 & 6.34 & 8.93 & 480 & 8.84 & 0.12 & 2.51 \\
50 & 5.02 & 6.85 & 9.18 & 180 & 8.44 & 0.18 & 1.59 \\}
\newcommand{\TabCV}{%
networks 1-1-1 & 14 & 0.126 & 0.168 & 0.135 & 0.140 & 0.134 & 0.159 \\
networks 1-3-1 & 26 & 0.119 & 0.163 & 0.127 & 0.143 & 0.128 & 0.164 \\
\textbf{networks 1-6-1} & 44 & 0.119 & 0.164 & 0.128 & 0.142 & 0.129 & 0.164 \\
networks 1-12-1 & 80 & 0.120 & 0.165 & 0.128 & 0.144 & 0.129 & 0.163 \\
\hline
\textbf{reduced model} & 6 & 0.128 & 0.170 & 0.137 & 0.139 & 0.136 & 0.159 \\
reduced model, $n_1=1$ & 5 & 0.136 & 0.176 & 0.144 & 0.146 & 0.143 & 0.165 \\
reduced model, $L(T)$ free & 9 & 0.126 & 0.168 & 0.135 & 0.138 & 0.133 & 0.157 \\
\hline
mean of retained weighings & --- & 0.094 & 0.162 & 0.104 & 0.158 & --- & --- \\}
\newcommand{\TabLam}{%
0 & 3.86 & 0.50--7.55 & 1.15 & 5 & 0.166 \\
0.01 & 1.34 & 0.50--1.43 & 1.17 & 32 & 0.165 \\
0.03 & 1.24 & 0.49--1.20 & 1.19 & 33 & 0.165 \\
0.1 & 0.71 & 0.49--0.77 & 1.12 & 177 & 0.164 \\
0.3 & 0.61 & 0.47--0.74 & 1.13 & 181 & 0.165 \\
1 & 0.43 & 0.40--0.52 & 1.13 & 445 & 0.169 \\}
\newcommand{\TabWeights}{%
\textbf{one pooled $\hat\sigma$} & 1.12 & 0.7 & 0.71 & 0.164 & 0.142 & 0.163 \\
one value per temperature & 1.16 & 3.7 & 0.70 & 0.172 & 0.148 & 0.165 \\
$s_{{\rm rep},j}$ at each point & 1.14 & 5.0 & 0.64 & 0.181 & 0.152 & 0.154 \\
one pooled $\hat\sigma$, weight decay $10^{-2}$ & 1.09 & 0.5 & 0.97 & 0.165 & 0.137 & 0.146 \\}
\newcommand{\TabParams}{%
$\kappa$ & ($10^{-3}$ min$^{-1}$) & 8.09 & 7.87 -- 8.46 \\
$E$ & (kJ mol$^{-1}$) & 33.9 & 32.8 -- 35.8 \\
$n_1$ &  & 1.143 & 1.105 -- 1.196 \\
$L_{50}$ & (g) & 0.67 & 0.54 -- 0.76 \\
$\theta_{\rm on}$ &  & 4.07 & 3.26 -- 7.00 \\
$\theta_2$ &  & 8.94 & 4.58 -- 15.29 \\}
\newcommand{\TabDerived}{%
$\tau=1/a$ & (min) & 348 & 199 & 124 & 43 \\
$U=G+L$ & (g) & 2.65 & 2.49 & 2.51 & 2.26 \\
$L$ & (g) & 0 & 0 & 0 & 0.67 \\}

%% file: body.tex
\title{Physics-constrained neural search for model discrimination and complexity reduction in seed imbibition}

\author{Luca Meacci}
\email{luca.meacci@unifi.it}
\affiliation{Department of Information Engineering, University of Florence, Florence, Italy}
\affiliation{Department of Mathematics and Computer Science ``Ulisse Dini,'' University of Florence, Florence, Italy}
\author{Angiolo Farina}
\affiliation{Department of Mathematics and Computer Science ``Ulisse Dini,'' University of Florence, Florence, Italy}
\author{Amin Aliyari}
\affiliation{School of Biosciences, University of Nottingham, Sutton Bonington, United Kingdom}
\author{Vincenzo di Bari}
\affiliation{School of Biosciences, University of Nottingham, Sutton Bonington, United Kingdom}
\author{Lorenzo Fusi}
\affiliation{Department of Mathematics and Computer Science ``Ulisse Dini,'' University of Florence, Florence, Italy}
\author{Duccio Fanelli}
\affiliation{Department of Physics and Astronomy, University of Florence, Florence, Italy}

\date{\today}

\begin{abstract}
A laboratory that follows seeds imbibing water records one number, the wet mass. Its evolution is
already complex: it rises and saturates, faster in a warmer bath, and in a hot bath it overshoots
and then decreases. Physical models of imbibition carry more structure than such a signal can
identify, while empirical laws fitted at each temperature do not say what the temperatures share.
We use machine learning to reduce the complexity of a model rather than to add to it. For rapeseed
batches at 4--50~$^\circ$C the rising data collapse, without fitting a model, onto one function of
a similarity variable that combines time and temperature through an Arrhenius clock. The wet mass
is then written as an uptake minus a mass loss, each a bounded increasing function of that
variable. Every such function is one minus the exponential of the integral of a nonnegative rate,
so that neural networks returning positive rates search every admissible kinetics. One model,
trained on all the measurements and tested on measurements it has not seen, returns for the uptake
a Weibull law with one exponent for all temperatures, $n_1=\nOne$, close to first order but
distinguishable from it on unseen data; a mass loss is required at 50~$^\circ$C alone. What
survives is a closed-form model with \nparWord\ constants for four temperatures, whose error on
unseen weighings is within \cvNNCFrel\% of that of the networks.
\end{abstract}

\maketitle

\section{Introduction}

The modelling of complex systems is concerned not only with many interacting degrees of freedom,
but also with the choice of a level of description that preserves testable structure without
assigning meaning to quantities the observations cannot resolve
\cite{Simon1962,Anderson1972,Ladyman2013,Ottino2004}. The choice is sharpest in partially observed
systems, where additional mechanistic detail produces practical non-identifiability rather than
additional testable content \cite{Gutenkunst2007,Raue2009,Miao2011,Transtrum2015}. The present
study is an instance of that problem, taken from seed science.

Seed imbibition is a simple experiment with a difficult modelling problem. Water uptake changes
the hydrated region, the pore space and the external shape of the seed, while temperature changes
the rate at which these processes develop \cite{Leopold1983,Louf2018}. The system considered here
is \textit{Brassica napus}, in which water entry is also spatially nonuniform
\cite{Munz2017,Lechowska2019}; related soft porous materials show nonmonotonic swelling and
delayed relaxation \cite{Aangenendt2020,ChenRaviChandar2022,Machado2026,Michailova2000}. The
experiment, however, records one number at a time: the wet mass of a batch.

\begin{figure}[htb]
 \centering
 \includegraphics[width=0.62\linewidth]{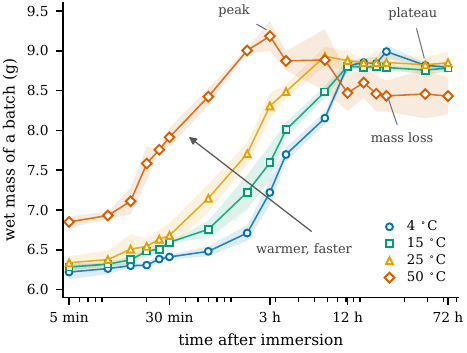}
 \caption{\textbf{The measurements.} Wet mass of a rapeseed batch after immersion in water at four
 bath temperatures: symbols are means over three replicate batches, bands one standard deviation
 between batches. The mass rises monotonically and reaches a plateau, faster in a warmer bath; at
 50~$^\circ$C it overshoots the plateau and then decreases.}
 \label{fig:data}
\end{figure}

Figure~\ref{fig:data} shows what that number does. The mass \textbf{rises monotonically} and
reaches a \textbf{plateau}; the warmer the bath, the \textbf{faster the rise}, by a factor of
about \factorSim\ between 4 and 50~$^\circ$C; and at 50~$^\circ$C the mass \textbf{overshoots} its
plateau, \textbf{peaks} at about three hours and then \textbf{decreases} for a day, in every
batch. Replicate batches differ: a single weighing scatters about the mean of its batches with a
standard deviation of \sigPool~g.

Such curves have been described at three levels. \textbf{Empirical laws} fit the uptake at one
temperature: first-order kinetics, the Peleg law \cite{Peleg1988}, and the Weibull law
\cite{Weibull1951}, which in food engineering is fitted at each temperature with its own scale and
shape \cite{MachadoOliveira1999,Marabi2003,Rafiq2015}. \textbf{Phenomenological readings} attach a
mechanism to the fitted numbers: the Weibull exponent, for instance, has been associated with
diffusion, external resistance or relaxation of the solid matrix \cite{Marabi2003}.
\textbf{Physical models} resolve the mechanism itself: Fickian diffusion with constant or
concentration-dependent diffusivity \cite{Crank1975,Hsu1983}, coupled diffusion and relaxation in
swelling polymers \cite{Peppas1994}, and free-boundary descriptions of the advancing hydration
front \cite{rubinshteuin1971stefan,fasano1983free,meuirmanov1992stefan,Meacci2026FreeBoundary}.

The \textbf{gap} addressed here lies between the detail a physical model can carry and the detail a
measurement can certify. A free-boundary description of the swelling seed involves fields (front
position, permeability, pore pressure) that a single aggregate signal cannot identify; an
empirical curve fitted to the same signal identifies its own parameters, but when it fails it does
not say whether the mechanism or the chosen formula is at fault. Model-discrimination methods ask
which dynamical descriptions are compatible with an observable
\cite{AtkinsonFedorov1975,Apgar2008,SkandaLebiedz2010,Lorenz2007,Hamadeh2013,Sontag2025}.
Scientific machine learning offers a further possibility: retain the known structure and leave only
a selected kinetic relation flexible
\cite{Rackauckas2020,Karniadakis2021,Masi2021,Flaschel2021,LinkaKuhl2023,Brunton2016,Chen2018};
related neural formulations exist for free-boundary and Stefan problems
\cite{WangPerdikaris2021,Long2026}. The question asked here is narrower than constitutive
identification or prediction at arbitrary temperature: \textit{which aspects of the observed
kinetics are shared across temperature, and when do the data require an additional response
mode?} Answering it is an exercise in \textbf{complexity reduction}.

\begin{figure}[htb]
 \centering
 \includegraphics[width=\linewidth]{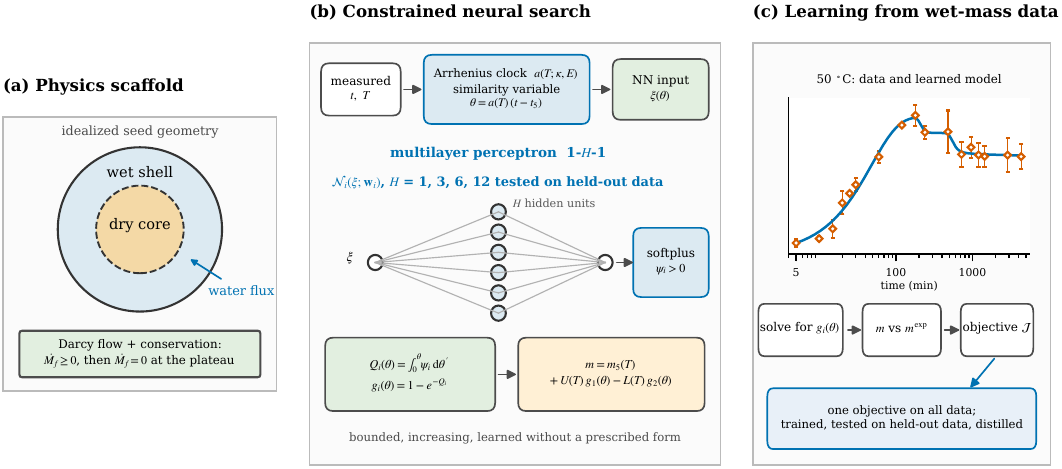}
 \caption{\textbf{Study map.} (a) The continuum description of the swelling seed, used only for
 the one implication a batch mass can test: the retained fluid mass increases until the plateau is
 reached. (b) The search. Time and temperature enter through the similarity variable, the only
 input of the networks; each network returns a positive rate, whose integral gives a bounded
 increasing function, and the phenomenological model combines the two functions into a predicted
 wet mass. (c) Learning from wet-mass data: one objective for all the data; the networks are
 trained on part of the data, tested on the rest, and what survives is distilled into closed
 form. The inset shows the learned model at 50~$^\circ$C.}
 \label{fig:pipeline}
\end{figure}

The argument proceeds in four steps (Fig.~\ref{fig:pipeline}), anticipated here and developed in
Secs.~\ref{sec:model}--\ref{sec:results}.

\textbf{(i) A similarity variable.} Each rising trajectory is normalized by its measured
endpoints, and time is rescaled by one factor per bath temperature. The four curves collapse onto
one. The factors are read off the measurements without fitting a model, and they follow an
Arrhenius law: time and temperature enter the response in one combination only.

\textbf{(ii) A phenomenological model.} The continuum description predicts a wet mass that never
decreases, which the 50~$^\circ$C batches violate. The mass is therefore written as the difference
of two bounded increasing contributions, an uptake and a mass loss, with shapes left unspecified
and amplitudes left free at every temperature.

\textbf{(iii) Machine learning for the kinetics.} Every bounded increasing function is one minus
the exponential of the integral of a nonnegative rate. A network returning a positive rate
therefore searches the admissible kinetics as a class, not a chosen formula. The two networks, one
per contribution, are trained together on all the measurements with one objective, in which a
small penalty on the mass loss lets the data decide where a loss is needed. The model is then
trained on part of the data and tested on the rest, so that a law that survives is a statement
about the data and not about a convenient choice of formula.

\textbf{(iv) The network returns the Weibull law.} For the uptake, the search returns the Weibull
law, with one exponent common to all temperatures, close to first order. Weibull introduced it in
1951 as the simplest admissible choice and stated that it had no theoretical basis
\cite{Weibull1951}; here it is what survives validation. Its argument is the similarity variable,
so that \textbf{one law in time and temperature} replaces the Weibull law fitted temperature by
temperature. A mass loss is required at 50~$^\circ$C alone; the data determine its
smallest admissible amplitude and bound its timing, but not its course.

In this sense a flexible function approximator \textbf{reduces the complexity of a model rather
than adding to it}: it represents every admissible law, and what survives validation is a formula
with \nparWord\ constants, each with a meaning, that predicts unseen data within a few per cent of
the networks. None of the steps is specific to seeds; they require a saturating observable and a
control parameter that rescales its clock.

\section{Data and phenomenological model}
\label{sec:model}

\subsection{The measurements}
\label{sec:data}

Three replicate batches were followed at bath temperatures $T=4$, 15, 25 and 50~$^\circ$C, at 18
times $t$ from 0 to 4320~min, in ultrapure water at a water-to-seed ratio of 4:1 by weight.
The first interval, from immersion to $t_5=5$~min, in which the mass rises from the dry value
$m_0$ to $m_5$ (Table~\ref{tab:data}), contains no intermediate weighing and is the most sensitive
to the draining and weighing protocol. The analysis therefore starts at $t_5$, takes $m_5$ as a
measured initial value and uses the 16 sampling times after it. The index $j=1,\dots,N$, with
$N=\Ntot$, runs over the pairs $(t_j,T_j)$ of sampling time and temperature. We write
$m^{\rm exp}_{j,b}$ for the wet mass of batch $b=1,2,3$ at $(t_j,T_j)$, $m^{\rm exp}_j$ for the
mean over the three batches and $s_{{\rm rep},j}$ for their standard deviation; $m$ without
superscript denotes the model. Each trajectory has four measured characteristics
(Table~\ref{tab:data}): $m_5(T)=m^{\rm exp}(t_5,T)$; the late level $m_{\rm late}(T)$, the mean
over $t\ge1440$~min; the largest mean mass $m_{\rm peak}(T)$; and the net gain
$G(T)=m_{\rm late}(T)-m_5(T)$.

\begin{table}[htb]
\caption{Measured characteristics of the four trajectories (means over three batches): dry mass
$m_0$; mass $m_5$ at $t_5=5$~min; largest mean mass $m_{\rm peak}$ and its time $t_{\rm peak}$;
late level $m_{\rm late}$; median over time of $s_{\rm rep}$; net gain $G=m_{\rm late}-m_5$. The
model takes $m_5$ and $G$ as measured inputs; $m_{\rm peak}$ enters only the normalization of
Eq.~\eqref{eq:normalize}.}
\label{tab:data}
\begin{ruledtabular}
\begin{tabular}{lrrrrrrr}
$T$ ($^\circ$C) & $m_0$ (g) & $m_5$ (g) & $m_{\rm peak}$ (g) & $t_{\rm peak}$ (min) & $m_{\rm late}$ (g) & $s_{\rm rep}$ (g) & $G$ (g) \\
\hline
\TabData
\end{tabular}
\end{ruledtabular}
\end{table}

\subsection{What the continuum description predicts, and where it fails}
\label{sec:baseline}

At continuum scale, capillary imbibition in a deformable seed is a free-boundary problem
\cite{rubinshteuin1971stefan,fasano1983free,meuirmanov1992stefan}; a formulation coupling Darcy
flow to the swelling skeleton is given in our companion work \cite{Meacci2026FreeBoundary}. For a
sample of conserved dry mass, the transport theorem with Darcy's law gives, for the retained fluid
mass $M_f$, $\dot M_f\ge0$ while capillary suction is directed inwards and $\dot M_f=0$ at
saturation: the wet mass rises and then reaches a plateau, whatever the permeability and the
geometry. At 4--25~$^\circ$C the largest mass of every batch exceeds its late level by at most
\overLowMax~g, an amount comparable with the differences between batches. At 50~$^\circ$C every
batch reaches its largest mass at 180~min and then loses \overFiftyList~g. The rise is admitted by
the continuum description; the decrease at 50~$^\circ$C is not.

\subsection{The similarity variable}
\label{sec:clock}

\begin{figure}[htb]
 \centering
 \includegraphics[width=0.62\linewidth]{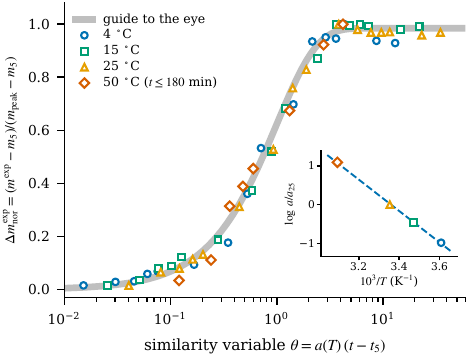}
 \caption{\textbf{The similarity variable.} $\Delta m^{\rm exp}_{\rm nor}$,
 Eq.~\eqref{eq:normalize}, on the rising data against $\theta=a(T)(t-t_5)$, with the shift factors
 obtained from level crossings: no curve is fitted. Inset: $\log a$ against $1/T$; the dashed line
 is the Arrhenius law of Eq.~\eqref{eq:clock}.}
 \label{fig:collapse}
\end{figure}

On the rising data (all times at 4--25~$^\circ$C, $t\le180$~min at 50~$^\circ$C) let
\begin{equation}
 \Delta m^{\rm exp}_{\rm nor}(t,T)=\frac{m^{\rm exp}(t,T)-m_5(T)}{m_{\rm peak}(T)-m_5(T)} .
 \label{eq:normalize}
\end{equation}
Our first \emph{ansatz} is that temperature acts as a clock: there is a shift factor $a(T)>0$ such
that $\Delta m^{\rm exp}_{\rm nor}$ depends on $t$ and $T$ only through $a(T)(t-t_5)$. The shift
factors are read off the data. For each level $\ell=0.10,0.15,\dots,0.90$ let $t_\ell(T)$ be the
first time at which $\Delta m^{\rm exp}_{\rm nor}(t,T)$, interpolated linearly between weighings,
reaches $\ell$; the ratio $a(T)/a(T')$ is the geometric mean over $\ell$ of
$[t_\ell(T')-t_5]/[t_\ell(T)-t_5]$. The four curves collapse (Fig.~\ref{fig:collapse}), and
$\log a$ is linear in $1/T$ within \residSim\%:
\begin{equation}
 \theta=a(T)\,(t-t_5),\qquad
 a(T)=\kappa\exp\!\left[-\frac{E}{R_g}\left(\frac{1}{T}-\frac{1}{T_{\rm ref}}\right)\right],
 \label{eq:clock}
\end{equation}
with $T$ in kelvin, $T_{\rm ref}=298.15$~K, $\kappa=a(T_{\rm ref})$, $E$ an effective activation
energy and $R_g$ the gas constant \cite{laidler1984development}. The slope gives
$E=\Esim$~kJ~mol$^{-1}$, about twice the value for the viscosity of water over the same range,
16.9~kJ~mol$^{-1}$: imbibition is not governed by the viscosity of the fluid alone.

\subsection{The phenomenological model}
\label{sec:phenomodel}

Our second \emph{ansatz} writes the mass as an uptake minus a mass loss, both functions of
$\theta$:
\begin{equation}
 m(t,T)=m_5(T)+U(T)\,g_1(\theta)-L(T)\,g_2(\theta),\qquad U(T)=G(T)+L(T),
 \label{eq:obs}
\end{equation}
with amplitudes $U,L\ge0$ in grams and shapes
\begin{equation}
 0\le g_i\le1,\qquad g_i(0)=0,\qquad g_i'\ge0,\qquad g_i\to1,\qquad i=1,2 .
 \label{eq:constraints}
\end{equation}
With $U=G+L$ the model returns the measured gain as $g_i\to1$. $L(T)$ is a free unknown at each
of the four temperatures: whether a mass loss is needed, and where, is left to the data.

\section{Learning the kinetics}
\label{sec:search}

\subsection{How a network searches every admissible kinetics}
\label{sec:family}

The constraints \eqref{eq:constraints} reduce to a single one if, as in Weibull's construction
\cite{Weibull1951},
\begin{equation}
 g_i(\theta)=1-e^{-Q_i(\theta)},\qquad Q_i(\theta)=\int_0^{\theta}\psi_i(\theta')\,{\rm d}\theta' ,
 \qquad \psi_i\ge0 :
 \label{eq:identity}
\end{equation}
every differentiable $g_i$ obeying \eqref{eq:constraints} has this form, with $Q_i=-\log(1-g_i)$,
and every nonnegative \textbf{rate} $\psi_i$ with divergent integral gives one. Since
$g_i'=\psi_i\,(1-g_i)$, $\psi_i$ is the rate of change of $g_i$ per unit of the change still to
come. The network represents the rate,
\begin{equation}
 \psi_i(\theta)=\operatorname{softplus}\!\left[\mathcal N_i(\xi;\mathbf w_i)\right]>0,\qquad
 \xi=\frac{\log(1+\theta)}{\log(1+\theta_{\max})} ,
 \label{eq:net}
\end{equation}
where $\mathcal N_i(\xi;\mathbf w_i)=\sum_{k=1}^{H}v_{ik}\tanh(\alpha_{ik}\xi+\beta_{ik})+c_i$ is a
network with one input, one hidden layer of $H$ units and one output (1-$H$-1), with weights
$\mathbf w_i=(\alpha_{ik},\beta_{ik},v_{ik},c_i)$, $3H+1$ in number and shared by all temperatures;
$\theta_{\max}=\kappa\,(4320-t_5)$ is the value of $\theta$ at the last weighing at $T_{\rm ref}$,
and $\xi$ maps the three decades of $\theta$ onto an interval of order one. For every value of the
weights, Eqs.~\eqref{eq:identity} and \eqref{eq:net} give an admissible kinetics.

\subsection{One model, all the data}
\label{sec:objective}

The weights $\mathbf w_1$, $\mathbf w_2$, the clock constants $\kappa$, $E$ and the four amplitudes
$L(T)$ minimize one objective over all $N=\Ntot$ means, the decrease at 50~$^\circ$C included:
\begin{equation}
 \begin{gathered}
 \mathcal J=\frac1{N\hat\sigma^2}\sum_{j=1}^{N}\left(m_j-m^{\rm exp}_j\right)^2
 +10^{-5}\left(\|\mathbf w_1\|^2+\|\mathbf w_2\|^2\right)+\lambda\sum_{T}L(T),\\
 \hat\sigma^2=\frac{1}{2N}\sum_{j=1}^{N}\sum_{b=1}^{3}\left(m^{\rm exp}_{j,b}-m^{\rm exp}_j\right)^2 ,
 \end{gathered}
 \label{eq:loss}
\end{equation}
where $m_j=m(t_j,T_j)$. The three terms are the misfit in units of the experimental scatter, a
weak regularization of the weights, and a penalty on the mass loss. The scatter
$\hat\sigma=\sigPool$~g is the standard deviation of a single weighing about the mean of its three
batches, pooled over the $N$ points; one pooled value is used because a standard deviation
estimated from three weighings is too uncertain to serve as a weight point by point
(Appendix~\ref{app:inv}).

The penalty is needed because the wet mass does not fix the decomposition \eqref{eq:obs}
uniquely: an earlier and larger loss, balanced by a larger uptake, leaves $m$ unchanged
(Appendix~\ref{app:inv}); without the penalty the learned amplitudes reach \LzeroMax~g. The term
$\lambda\sum_TL(T)$ selects, among such equivalent decompositions, the one with the smallest loss:
\textbf{the loss is used only where the data require it}. We take $\lambda=\lamVal$~g$^{-1}$, the
smallest value at which the amplitudes learned from different pairs of batches agree within
\spFoldsSel~g; the test error of Sec.~\ref{sec:validation} does not depend on $\lambda$
(Appendix~\ref{app:inv}).

The networks have $H=6$, with \nwSix\ weights each and \nparSix\ unknowns in all; $H=1$, 3 and 12
are trained alongside. Of twelve random initializations, the one with the lowest $\mathcal J$ is
the \textbf{reference}: the data-driven solution $g_1$, $g_2$, $\kappa$, $E$, $L(T)$ from which
every reduction below starts and with which it is compared (Appendix~\ref{app:refred}).

\subsection{The network learns: trained on part of the data, tested on the rest}
\label{sec:validation}

A network that has learned the kinetics predicts weighings on which it was not trained; one that
has memorized the data does not \cite{Varma2006,Cawley2010,Roberts2017}. The \Nweigh\ weighings
$(j,b)$ are therefore split into a training set $S_{\rm tr}$ and a test set $S_{\rm te}$. The whole
model is trained afresh on $S_{\rm tr}$, with $m_5$, $G$ and $\hat\sigma$ recomputed from
$S_{\rm tr}$, and its error on a set $S$ is the root-mean-square difference, in grams, between the
model and the individual weighings of $S$,
\begin{equation}
 \varepsilon(S)=\Big[\frac{1}{|S|}\sum_{(j,b)\in S}\big(m_j-m^{\rm exp}_{j,b}\big)^2\Big]^{1/2} ;
 \label{eq:rmse}
\end{equation}
$\varepsilon(S_{\rm tr})$ is the training error and $\varepsilon(S_{\rm te})$ the test error. Three
choices of $S_{\rm te}$ probe three kinds of prediction:
\begin{itemize}
\item[(a)] a new batch: $S_{\rm te}=\{(j,b):b=b^*\}$, $b^*=1,2,3$; three folds of 64 test weighings;
\item[(b)] missing weighings: $S_{\rm te}$ consists of \NrandTest\ weighings drawn at random
across batches, temperatures and times; \NranSplits\ independent splits;
\item[(c)] a missing sampling time: $S_{\rm te}=\{(j,b):t_j=t^*\}$ for each of the 16 times $t^*$;
\NtimeFolds\ folds of 12 test weighings.
\end{itemize}
The test error of a scheme is Eq.~\eqref{eq:rmse} evaluated on the union of its test sets. Two
reference values set its scale. A model equal to the exact mean has, on unseen weighings, the
expected error $\hat\sigma=\sigPool$~g; no model can have a smaller one. In scheme (a), the mean
of the two retained batches, which uses no model, has test error $\hat\sigma\sqrt{3/2}=\teBatREF$~g.

Table~\ref{tab:cv} gives the results. In scheme (b) the 1-6-1 networks reach
$\varepsilon(S_{\rm te})=\teRanNNsix$~g, \relSig\% above $\hat\sigma$. In scheme (a) they reach
\teBatNNsix~g: a new batch is predicted as well as the other two batches predict it. In scheme (c),
where a sampling time is missing at every temperature, they reach \teTimNNsix~g. The four widths
give test errors within \widthSpread~mg of each other, and training to convergence raises the test
error by at most a few milligrams (Appendix~\ref{app:tt}). \textbf{The networks learn the
kinetics; they do not memorize the measurements.}

\begin{table}[htb]
\caption{Training error $\varepsilon(S_{\rm tr})$ and test error $\varepsilon(S_{\rm te})$,
Eq.~\eqref{eq:rmse}, in grams, for the three choices of test set of Sec.~\ref{sec:validation}.
Every model is trained afresh in every fold; $p$ is its number of unknowns. Networks:
Sec.~\ref{sec:objective}. Reduced model and its two variants: Sec.~\ref{sec:closed}. Last row:
the mean of the retained weighings at the same $(t_j,T_j)$, which uses no model (undefined in (c),
where no weighing at $t^*$ is retained). Reference value for the test error:
$\hat\sigma=\sigPool$~g. A training error below $\hat\sigma$ is not a sign of overfitting
(Appendix~\ref{app:tt}).}
\label{tab:cv}
\footnotesize
\begin{ruledtabular}
\begin{tabular}{lrcccccc}
 & & \multicolumn{2}{c}{(a) new batch} & \multicolumn{2}{c}{(b) missing weighings} & \multicolumn{2}{c}{(c) missing time} \\
\cline{3-4}\cline{5-6}\cline{7-8}
model & $p$ & training & test & training & test & training & test \\
\hline
\TabCV
\end{tabular}
\end{ruledtabular}
\end{table}

\subsection{The network returns the Weibull law}
\label{sec:returns}

\begin{figure}[htb]
 \centering
 \includegraphics[width=\linewidth]{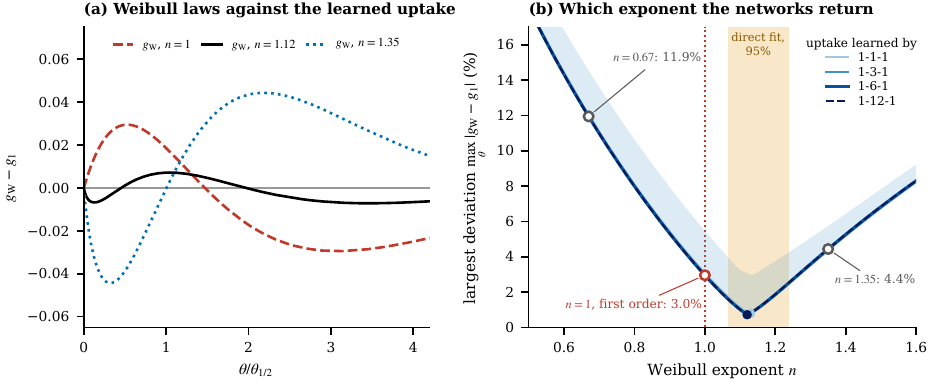}
 \caption{\textbf{The network returns the Weibull law.} Distances between functions,
 Eq.~\eqref{eq:D}; no data enter. (a) $g_{\rm W}-g_1$ for the closest Weibull laws of three
 exponents and the median uptake of the twelve 1-6-1 networks, against $\theta/\theta_{1/2}$, with
 $g_1(\theta_{1/2})=1/2$. (b) $D(n)$ for networks of 1, 3, 6 and 12 hidden units: median (lines) and
 range (bands) over twelve initializations; dots mark the minima, the exponent each network has
 learned. Shaded: the 95\% interval of $n_1$ fitted to the data.}
 \label{fig:distil}
\end{figure}

The networks return a function $g_1(\theta)$, not a formula. To identify it, functions are
compared with functions, without reference to the data: for each exponent $n$, the scale
$\lambda_{\rm W}$ of a Weibull law is chosen to minimize its largest distance from $g_1$ over the
range of $\theta$ covered by the data,
\begin{equation}
 D(n)=\min_{\lambda_{\rm W}}\ \max_{\theta}\ \left|1-e^{-(\theta/\lambda_{\rm W})^n}-g_1(\theta)\right| ,
 \label{eq:D}
\end{equation}
and the exponent that minimizes $D$ identifies the Weibull law closest to what the network has
learned. For all 48 networks trained on all the data (four widths, twelve initializations each)
the minimum lies at $n=\nSupMin$--$\nSupMax$ (Fig.~\ref{fig:distil}). There $D=\supBest\%$ of the
uptake for the 1-6-1 networks, against \supOne\% at $n=1$, \supHigh\% at $n=1.35$ and \supDiff\% at
$n=0.67$. The networks also return one clock, $E=\EnnMin$--$\EnnMax$~kJ~mol$^{-1}$, close to the
model-free value of Sec.~\ref{sec:clock}.

The first reduction therefore replaces $g_1$ by the Weibull law
\begin{equation}
 g_{\rm W1}(\theta)=1-e^{-\theta^{\,n_1}} ,
 \label{eq:weibull}
\end{equation}
with $\lambda_{\rm W}$ absorbed in $\kappa$. Equation~\eqref{eq:D} identifies the form; the
constants are not taken from it. $\kappa$, $E$ and $n_1$ are fitted to the measured means by
minimizing Eq.~\eqref{eq:loss}, together with the constants of the mass loss
(Sec.~\ref{sec:closed}); the fit gives $n_1=\nOneLong$, with 95\% interval \nOneLoNF--\nOneHiNF.
In the variables of the experiment, Eq.~\eqref{eq:weibull} reads
\begin{equation}
 g_{\rm W1}=1-\exp\!\left\{-\left[\frac{t-t_5}{\tau(T)}\right]^{n_1}\right\},\qquad
 \tau(T)=\frac{1}{a(T)}=\frac{1}{\kappa}\exp\!\left[\frac{E}{R_g}\left(\frac{1}{T}-\frac{1}{T_{\rm ref}}\right)\right] .
 \label{eq:weibullt}
\end{equation}

Each constant has a definite content (Table~\ref{tab:use}). The \textbf{scale} $\tau(T)$ is the
time at which $g_{\rm W1}=1-e^{-1}$, that is, 63\% of the uptake: \tauFour, \tauFifteen,
\tauTwentyFive\ and \tauFifty~min at 4, 15, 25 and 50~$^\circ$C. Its dependence on temperature is
not fitted temperature by temperature: it is the clock of Eq.~\eqref{eq:clock}. The \textbf{shape}
is one exponent for all temperatures: temperature changes the rate of imbibition, not the form of
its course. By Eq.~\eqref{eq:identity}, the rate per unit of remaining uptake is
$\psi_1=n_1\theta^{\,n_1-1}$, constant for first order and here slowly increasing, as
$\theta^{\,\nOneMinusOne}$. The exponent is \textbf{close to one but not one}: with $n_1=1$ the
test error of the reduced model increases in every scheme, by \cvBatFO, \cvRanFO\ and \cvTimFO~mg
(95\% intervals: \cvBatFOLo\ to \cvBatFOHi, \cvRanFOLo\ to \cvRanFOHi\ and \cvTimFOLo\ to
\cvTimFOHi~mg). In the classification of Ref.~\cite{Marabi2003} for spheres, $n=0.67$ corresponds
to Fickian diffusion, $n=1$ to a first-order process limited by an external resistance and
$n\approx1.2$ to relaxation of the solid matrix; the 95\% interval of $n_1$ excludes the first two
values and contains the third.

\begin{table}[htb]
\caption{The Weibull law in food engineering and in this work, for an uptake measured at $K$
temperatures. $\alpha$, $\beta$: scale and shape of the Weibull law in its usual form; the other
symbols as in the text.}
\label{tab:use}
\footnotesize
\begin{ruledtabular}
\begin{tabular}{>{\raggedright\arraybackslash}p{0.22\linewidth}>{\raggedright\arraybackslash}p{0.34\linewidth}>{\raggedright\arraybackslash}p{0.36\linewidth}}
 & usual practice & this work \\
\hline
form & $1-\exp[-(t/\alpha)^\beta]$ \cite{Weibull1951,Cunha1998,Marabi2003} & $1-\exp(-\theta^{\,n_1})$, Eq.~\eqref{eq:weibull} \\
argument & time at one temperature \cite{MachadoOliveira1999,Marabi2003} & $\theta=a(T)(t-t_5)$, all temperatures, Eq.~\eqref{eq:clock} \\
scale & $\alpha(T)$, fitted at each $T$ \cite{Marabi2003,Rafiq2015} & $\tau(T)=1/a(T)$, Eq.~\eqref{eq:weibullt} \\
temperature dependence & law fitted to $\alpha(T)$ afterwards \cite{Rafiq2015} & shift factors measured without a model, Sec.~\ref{sec:clock} \\
shape & $\beta(T)$, fitted at each $T$ \cite{Marabi2003} & one $n_1$ for all $T$ \\
constants & $2K$, plus those of the law for $\alpha(T)$ & 3: $\kappa$, $E$, $n_1$ \\
meaning of the shape & $\beta$ compared with 0.67 (diffusion), 1 (external resistance), $\approx1.2$ (relaxation) \cite{Marabi2003} & the same comparison for $n_1$, with its 95\% interval \\
location parameter & usually zero & onset $\theta_{\rm on}$ of the mass loss, Eq.~\eqref{eq:states} \\
choice of the form & assumed & selected by a search over all $g$ obeying Eq.~\eqref{eq:constraints}, tested on unseen data \\
\end{tabular}
\end{ruledtabular}
\end{table}

\FloatBarrier
\section{Mass loss}
\label{sec:loss}

\subsection{Where the mass loss appears}
\label{sec:where}

The 48 networks return $L=\LnetFiftyMin$--$\LnetFiftyMax$~g at 50~$^\circ$C, at most
\LnetMidMax~g at 15 and 25~$^\circ$C, and \LnetFourMin--\LnetFourMax~g at 4~$^\circ$C, where the
weighings at 1440~min are heavier than those at 1200~min in every batch. Whether the small
amplitudes are needed is decided on unseen data: with $L(T)$ free at every temperature, the test
error of the reduced model changes by at most \cvLallMax~mg in the three schemes
(Table~\ref{tab:cv}), and the 95\% upper bounds of $L$ at 4, 15 and 25~$^\circ$C are \LallHiFour,
\LallHiFifteen\ and \LallHiTwentyFive~g. \textbf{A mass loss is required at 50~$^\circ$C alone},
and the second reduction sets $L=0$ at 4--25~$^\circ$C. At 50~$^\circ$C the uptake amplitude is
then $U=G+L=\UFifty$~g, close to the measured $m_{\rm peak}-m_5=\ampFifty$~g.

\subsection{When the mass is lost}
\label{sec:returnsloss}

Three facts are read directly from the weighings at 50~$^\circ$C: every batch reaches its largest
mass at 180~min; every batch is lighter at 240~min, by \dropMin--\dropMax~g; and the mean at
720~min is within \mSevenTwentyDiff~g of the late level. On the means, the loss has begun by
240~min and is complete by 720~min.

The networks add one statement and leave one question open. For $H\ge3$ the learned $g_2$ stays
below 1\% until \tGtwoWideMin--\tGtwoWideMax~min and reaches \gTwoAtTwoForty\%,
\gTwoAtFourEighty\% and \gTwoAtSevenTwenty\% at 240, 480 and 720~min (with $H=1$ the onset is
smoother, with 1\% reached at \tGtwoOneNet~min). This onset at the peak is that of the smallest
loss compatible with the data, the decomposition selected by $\lambda$: a larger loss,
\LearlyMin--\LearlyMax~g, starting within \tEarlyMax~min and balanced by a larger uptake, gives the
same test error (Appendix~\ref{app:inv}). The open question is the course of the loss between the
peak and 720~min: the interval contains two weighings, at 240 and 480~min, and at 480~min the
batches span \spreadFourEighty~g, more than $L$. The learned $g_2$ follows the means at these two
times and is not a single Weibull law (the closest three-parameter form deviates from it by
\supLossFree\%).

The third reduction therefore replaces $g_2$ by the simplest admissible law with an onset,
Weibull's three-parameter form \cite{Weibull1951} with unit exponent,
\begin{equation}
 g_{\rm W2}(\theta)=1-\exp\!\left(-\frac{\theta-\theta_{\rm on}}{\theta_2}\right)\ \ \text{for }\theta>\theta_{\rm on},\qquad g_{\rm W2}=0\ \ \text{otherwise},
 \label{eq:states}
\end{equation}
where the location parameter $\theta_{\rm on}$ is the onset of the loss and $\theta_2$ its scale.
Fitted to the data (Sec.~\ref{sec:closed}), it gives at 50~$^\circ$C the onset
$t_{\rm on}=t_5+\theta_{\rm on}/a=\tOn$~min (68\% interval: \tOnLo--\tOnHi) and the characteristic
time $\tau_2=\theta_2/a=\tauTwo$~min (68\% interval: \tauTwoLo--\tauTwoHi), the least determined
constants of the model. The result is a gradual decrease from the peak to the plateau
(Fig.~\ref{fig:model}) that stays within the range of the three batches at every time after the
peak and has the same test error as the networks within \cvNNCFmax~mg (Table~\ref{tab:cv}). The
natural reading of the loss is the release of seed material into the water, accelerated by
temperature \cite{Aliyari2026}; the late level at 50~$^\circ$C is 0.4~g lower than at
4--25~$^\circ$C.

\FloatBarrier
\section{A ready-to-use reduced model}
\label{sec:results}

\subsection{Six constants for four temperatures}
\label{sec:closed}

\begin{figure}[htb]
 \centering
 \includegraphics[width=\linewidth]{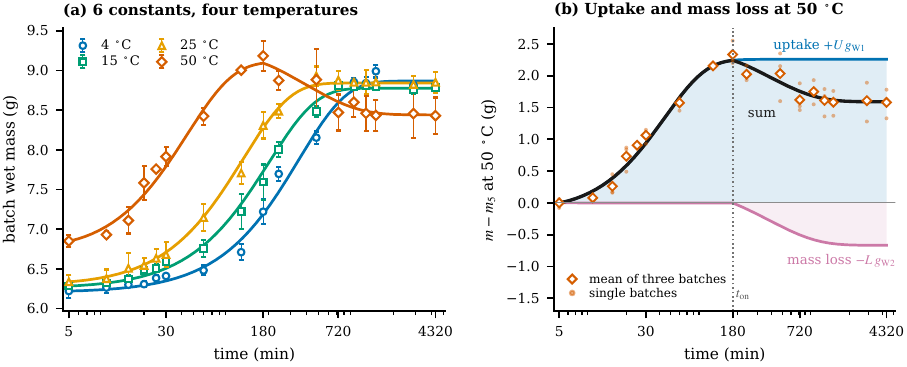}
 \caption{\textbf{The reduced model.} (a) Measured means with one standard deviation between
 batches, and the model with the \npar\ constants of Table~\ref{tab:params} and the measured $m_5$
 and $G$. (b) At 50~$^\circ$C, $m-m_5$ split into the uptake $U g_{\rm W1}$ and the mass loss
 $L g_{\rm W2}$; diamonds: means; dots: single batches; vertical line: $t_{\rm on}$.}
 \label{fig:model}
\end{figure}

The reduced model is the reference with the three reductions: $g_1\to g_{\rm W1}$,
Eq.~\eqref{eq:weibull}; $L\to0$ at 4--25~$^\circ$C; $g_2\to g_{\rm W2}$, Eq.~\eqref{eq:states}. It has
\begin{equation}
 \underbrace{2}_{\kappa,\,E}+\underbrace{1}_{n_1}+\underbrace{1}_{L_{50}}
 +\underbrace{2}_{\theta_{\rm on},\,\theta_2}=\npar
 \label{eq:count}
\end{equation}
constants for all four temperatures, in addition to the measured $m_5(T)$ and $G(T)$ of each
trajectory. The six constants are fitted together to the \Ntot\ measured means by minimizing
Eq.~\eqref{eq:loss} with $g_1$, $g_2$ replaced by $g_{\rm W1}$, $g_{\rm W2}$; the Weibull laws
closest to the reference provide only the starting values. Table~\ref{tab:params} gives the
constants; Fig.~\ref{fig:model} compares the model with the measurements, and
Fig.~\ref{fig:refred} with the reference. The root-mean-square difference from the measured means,
$\bar\varepsilon=[N^{-1}\sum_j(m_j-m^{\rm exp}_j)^2]^{1/2}$, is \rmseFin~g, against \rmseNet~g for
the reference with its \nparSix\ unknowns. On unseen weighings (Table~\ref{tab:cv}) the two
predict within \cvNNCFmax~mg of each other: the networks do better in scheme (a), the reduced
model in schemes (b) and (c).

\begin{table}[htb]
\caption{The reduced model. Top: the six constants fitted to the measured means, with 68\%
intervals from a parametric bootstrap (Appendix~\ref{app:num}); $\theta_{\rm on}$ and $\theta_2$
are in the unit of $\theta$ fixed by Eq.~\eqref{eq:weibull}. Bottom: at each temperature, the time
$\tau=1/a$ to 63\% of the uptake and the amplitudes $U$ and $L$.}
\label{tab:params}
\begin{ruledtabular}
\begin{tabular}{llrl}
\multicolumn{2}{l}{\emph{constants}} & value & 68\% interval \\
\hline
\TabParams
\end{tabular}
\end{ruledtabular}
\vspace{6pt}
\begin{ruledtabular}
\begin{tabular}{llrrrr}
\multicolumn{2}{l}{\emph{derived}} & 4~$^\circ$C & 15~$^\circ$C & 25~$^\circ$C & 50~$^\circ$C \\
\hline
\TabDerived
\end{tabular}
\end{ruledtabular}
\end{table}

\subsection{The model as a working formula}
\label{sec:formula}

Figure~\ref{fig:recipe} collects the model in the form in which it is applied. At a temperature
between those measured, the clock of Eq.~\eqref{eq:clock} gives $\theta$ and the two shapes apply
unchanged; $m_5$ and $G$ are measured on the batch at hand. The amplitude $L$ is known at
50~$^\circ$C and bounded at 4--25~$^\circ$C (Sec.~\ref{sec:where}); between 25 and 50~$^\circ$C
these data do not determine it.

\begin{figure}[htb]
\centering
\fbox{\begin{minipage}{0.96\linewidth}
\small
\textbf{Wet mass of a rapeseed batch for $t\ge t_5=5$~min}
\par\medskip
\begin{tabular}{@{}>{\raggedright\arraybackslash}p{0.21\linewidth}>{\raggedright\arraybackslash}p{0.74\linewidth}@{}}
\emph{1. Clock} & $a(T)=\kappa\exp\!\left[-\dfrac{E}{R_g}\left(\dfrac1T-\dfrac1{T_{\rm ref}}\right)\right]$,\quad
$\theta=a(T)\,(t-t_5)$\\[10pt]
\emph{2. Uptake} & $g_{\rm W1}=1-\exp\!\left(-\theta^{\,n_1}\right)$\\[6pt]
\emph{3. Mass loss} & $g_{\rm W2}=1-\exp\!\left[-(\theta-\theta_{\rm on})/\theta_2\right]$ for $\theta>\theta_{\rm on}$, $g_{\rm W2}=0$ otherwise\\[6pt]
\emph{4. Wet mass} & $m(t,T)=m_5(T)+U(T)\,g_{\rm W1}-L(T)\,g_{\rm W2}$,\quad $U=G+L$\\
\end{tabular}
\par\medskip\hrule\medskip
\begin{tabular}{@{}>{\raggedright\arraybackslash}p{0.21\linewidth}>{\raggedright\arraybackslash}p{0.74\linewidth}@{}}
\emph{Constants} & $\kappa=\kappaVal\times10^{-3}$~min$^{-1}$,\ $E=\Efin$~kJ~mol$^{-1}$,\ $n_1=\nOneLong$,\
$\theta_{\rm on}=\thetaOn$,\ $\theta_2=\thetaTwo$;\ \ $T_{\rm ref}=298.15$~K,\ $R_g=8.314\times10^{-3}$~kJ~mol$^{-1}$~K$^{-1}$\\[6pt]
\emph{Inputs} & $m_5$ and $G=m_{\rm late}-m_5$, measured;\ \ $L=0$ at 4--25~$^\circ$C,\ $L=\LFifty$~g at 50~$^\circ$C\\[6pt]
\emph{Times} & 63\% of the uptake at $t_5+1/a(T)$;\ onset of the loss at $t_5+\theta_{\rm on}/a(T)$\\
\end{tabular}
\end{minipage}}
\caption{\textbf{The model as a working formula.} Constants from Table~\ref{tab:params}; $T$ in
kelvin.}
\label{fig:recipe}
\end{figure}

\FloatBarrier
\section{Conclusion}

The question posed at the outset has a definite answer for these measurements.

\begin{enumerate}
\item \textbf{What is shared across temperature is the shape of the uptake.} On the data where the
mass rises, the four temperatures are one function of a similarity variable; its shift factors
are measured rather than modelled, and follow an Arrhenius law over 4--50~$^\circ$C.
\item \textbf{The uptake is an effective Weibull law of time and temperature together.} Networks
of every width, trained on all the data and tested on data they have not seen, return the
Weibull law, with one exponent for all temperatures, close to first order and distinguishable
from it on unseen data. The law proposed by Weibull in 1951 for its simplicity, and used since
then temperature by temperature, is recovered from the measurements, with the similarity variable
as its argument.
\item \textbf{The data require a mass loss at 50~$^\circ$C alone.} Its amplitude, free at every
temperature and penalized, is switched on by the data at 50~$^\circ$C only; in its smallest form
the loss starts at the peak, and its course towards the plateau is the least determined element
of the model.
\end{enumerate}

\textbf{Machine learning as an instrument of complexity reduction.} The method has four steps: a
similarity variable read off the data; the admissible kinetics written through one nonnegative
rate; a neural search of that whole class, with \textbf{one objective for all the data}, trained
and tested on separate data; and the distillation of what the network returns into closed form.
In this use the network is \textbf{not the model but a probe}: it represents every admissible law,
so that the law which survives validation is selected by the data and not by the modeller. That a
search free to return any admissible kinetics comes back with Weibull's law is the result, not a
limitation: it certifies from the data a law used for decades, and gives its scale and shape a
definite content. The networks, with \nparSix\ unknowns, are replaced by a formula with
\textbf{\nparWord\ constants, each with a meaning}, that predicts unseen data within a few per
cent of them. The resulting law is \textbf{effective}, not fundamental: it fixes what the
measurement certifies and leaves to physical models what it does not. Nothing in the steps is
specific to seeds: they apply to any saturating observable whose control parameters collapse into
a similarity variable.

The results open three directions. Measurements between 25 and 50~$^\circ$C, with a denser
sampling after the peak, would show how the mass loss grows with temperature, whether it keeps the
clock of the uptake, and what its course is. The composition of the bath would identify the
released material. Spatially resolved hydration measurements
\cite{Munz2017,Manz2005,Kikuchi2006} would show where inside the seed the two processes take
place, the point at which the description obtained here connects to a free-boundary formulation
\cite{Meacci2026FreeBoundary}.

\begin{acknowledgments}
L.M., L.F. and A.F. acknowledge the support of the Gruppo Nazionale per la Fisica Matematica
(GNFM) of the Istituto Nazionale di Alta Matematica (INdAM), Italy.
\end{acknowledgments}

\section*{Data availability}
The data that support the findings of this article, together with the code that regenerates every
figure, table and number, are available from the corresponding author upon reasonable request.

\appendix

\section{Numerical details}
\label{app:num}

\emph{Networks.} The unknowns $\kappa$, $E$ and $L(T)$ are kept positive by a softplus
transformation. The weights are initialized as in Ref.~\cite{Glorot2010}, and the output bias of
the loss network is initialized so that the loss starts small. $Q_i$ is obtained by trapezoidal
quadrature on a logarithmic grid of $\theta$. The objective $\mathcal J$ is minimized on the full
data set by Adam \cite{Kingma2015}, with step $0.015$ and 20\,000 iterations, in JAX
\cite{Bradbury2018}. Twelve random initializations are run on all the data and three in every fold
of Sec.~\ref{sec:validation}; the run with the lowest $\mathcal J$ is retained.

\emph{Validation.} In every fold $m_5$, $G$ and $\hat\sigma$ are recomputed from the retained
weighings and every model of Table~\ref{tab:cv} is fitted afresh. No width and no setting is
selected on test data. The difference between the test errors of two models is evaluated on the
same test weighings; its 95\% interval is obtained by bootstrap resampling of the folds
\cite{Efron1993}.

\emph{Reduced model and uncertainties.} The six constants minimize Eq.~\eqref{eq:loss} by bounded
least squares, started from the closed forms closest to the best 1-6-1 networks. Their intervals
come from a parametric bootstrap \cite{Efron1993}: \Bboot\ synthetic data sets are generated from
the fitted model with the scatter of its residuals, $m_5$ and $G$ are redrawn within the standard
errors of their measured means, and each data set is refitted; the 68\% and 95\% intervals are the
corresponding percentiles.

\emph{Shift factors.} At 50~$^\circ$C the shift factor of Sec.~\ref{sec:clock} is \aFiftyCutOne\ or
\aFiftyCutTwo~$\times10^{-3}$~min$^{-1}$ when the rising data are cut at 120 or at 180~min.

\section{The penalty on the mass loss and the weights}
\label{app:inv}

\emph{Invariance.} For any $\mu\ge1$ the replacement
\begin{equation}
 L\to\mu L,\qquad g_2\to\Big(1-\frac1\mu\Big)\,g_1+\frac{g_2}{\mu}
 \label{eq:inv}
\end{equation}
preserves Eq.~\eqref{eq:constraints} and leaves $m$ in Eq.~\eqref{eq:obs} unchanged, since
$(G+\mu L)g_1-\mu L[(1-1/\mu)g_1+g_2/\mu]=(G+L)g_1-Lg_2$. The first term of $\mathcal J$ cannot
distinguish these decompositions; the penalty $\lambda\sum_TL(T)$ selects the one with the smallest
$L$.

\emph{Choice of $\lambda$.} Table~\ref{tab:lam} shows the dependence on $\lambda$. At $\lambda=0$
the amplitude is arbitrary; at $\lambda=0.01$--0.03 the loss is \LearlyMin--\LearlyMax~g and starts
within \tEarlyMax~min; from $\lambda=\lamLate$~g$^{-1}$ on it starts at the peak. The amplitudes
learned on the three pairs of batches of scheme (a) spread by \spFoldsSel~g at $\lambda=0.1$,
against \spFoldsLow~g at $\lambda=0.03$. For every
$\lambda\le0.3$~g$^{-1}$ the test error is \lamTestMin--\lamTestMax~g and the closest Weibull
exponent lies between \lamNmin\ and \lamNmax: $\lambda$ selects the decomposition, not the
prediction.

\begin{table}[htb]
\caption{Dependence of the learned model on $\lambda$ in Eq.~\eqref{eq:loss} (1-6-1 networks).
$L_{50}$: amplitude of the loss at 50~$^\circ$C learned on all the data, and its range over the
three pairs of training batches of scheme (a); $n_1$: exponent of the closest Weibull law,
Eq.~\eqref{eq:D}; $t_{1\%}$: time at which $g_2=0.01$ at 50~$^\circ$C; $\varepsilon(S_{\rm te})$:
test error of scheme (a), Eq.~\eqref{eq:rmse}.}
\label{tab:lam}
\begin{ruledtabular}
\begin{tabular}{lccccc}
$\lambda$ (g$^{-1}$) & $L_{50}$ (g) & $L_{50}$, scheme (a) (g) & $n_1$ & $t_{1\%}$ (min) & $\varepsilon(S_{\rm te})$ (g) \\
\hline
\TabLam
\end{tabular}
\end{ruledtabular}
\end{table}

\emph{Weights and regularization.} Table~\ref{tab:weights} compares the pooled $\hat\sigma$ of
Eq.~\eqref{eq:loss} with two alternatives: one value per temperature (from \sRepFour~g at
4~$^\circ$C to \sRepFifty~g at 50~$^\circ$C), and the standard deviation $s_{{\rm rep},j}$ of the
three weighings at each point. The latter has two degrees of freedom: as a weight it would make
the point at 480~min and 50~$^\circ$C ($s_{\rm rep}=\sRepFourEighty$~g) count \wRatio\ times less
than the point at 240~min. With either alternative the closest Weibull exponent remains
\wAltNmin--\wAltNmax\ and the loss at 50~$^\circ$C \wAltLmin--\wAltLmax~g, while the learned
uptake lies farther from a Weibull law and the test errors of schemes (a) and (b) are larger. The
last row shows a strong weight decay, the usual regularization of a network. It is not used
because it drives the rate $\psi_1$ towards a constant, that is, towards first order, the
hypothesis under test: the exponent falls to \nWD, and the loss returned is one of the earlier and
larger ones of Eq.~\eqref{eq:inv}.

\begin{table}[htb]
\caption{Dependence of the learned model on the weights and on the weight decay in
Eq.~\eqref{eq:loss} (1-6-1 networks, $\lambda=\lamVal$~g$^{-1}$). $n_1$ and $D$: exponent of the
closest Weibull law and its distance from the learned uptake, Eq.~\eqref{eq:D}; $L_{50}$:
amplitude of the loss at 50~$^\circ$C; last three columns: test error $\varepsilon(S_{\rm te})$ in
schemes (a)--(c). In bold, the choice used.}
\label{tab:weights}
\begin{ruledtabular}
\begin{tabular}{lcccccc}
weights & $n_1$ & $D$ (\%) & $L_{50}$ (g) & (a) (g) & (b) (g) & (c) (g) \\
\hline
\TabWeights
\end{tabular}
\end{ruledtabular}
\end{table}

\FloatBarrier
\section{The reference and the reduced model}
\label{app:refred}

\begin{figure}[tbp]
 \centering
 \includegraphics[width=0.86\linewidth]{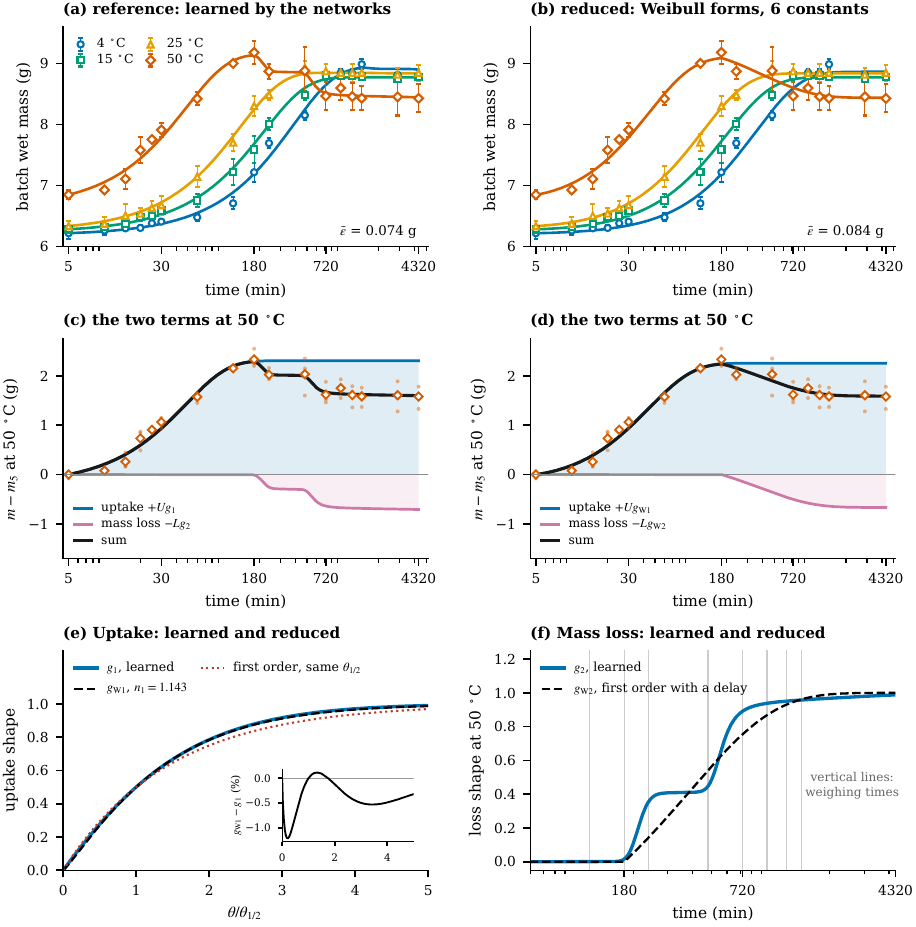}
 \caption{\textbf{The reference and the reduced model}: what the three reductions change. Left:
 learned by the 1-6-1 networks (Sec.~\ref{sec:objective}). Right: reduced model,
 Table~\ref{tab:params}. (a, b) Means with one standard deviation between batches;
 $\bar\varepsilon$ over the \Ntot\ means. (c, d) At 50~$^\circ$C, uptake and loss terms of
 Eq.~\eqref{eq:obs} and their sum; diamonds: means, dots: batches. (e) $g_1$ and $g_{\rm W1}$
 against $\theta/\theta_{1/2}$, with first order of equal $\theta_{1/2}$; inset:
 $g_{\rm W1}-g_1$. (f) $g_2$ and $g_{\rm W2}$ against time at 50~$^\circ$C; vertical lines:
 weighing times.}
 \label{fig:refred}
\end{figure}

Figure~\ref{fig:refred} sets the reference beside the reduced model. The Weibull law of the
reduced model departs from the learned uptake by at most \devRefRed\% (panel e). The reference
keeps the 4~$^\circ$C weighings at 1440~min as a small mass loss, which the reduced model sets to
zero (Sec.~\ref{sec:where}). At 50~$^\circ$C the two place the same amplitude of the loss after the
same onset and differ in its course (panels c, d, f): the reference follows the means at 240 and
480~min, the reduced model decreases gradually, and both stay within the spread between batches.

\FloatBarrier
\section{Training and test error during training}
\label{app:tt}

\begin{figure}[htb]
 \centering
 \includegraphics[width=0.62\linewidth]{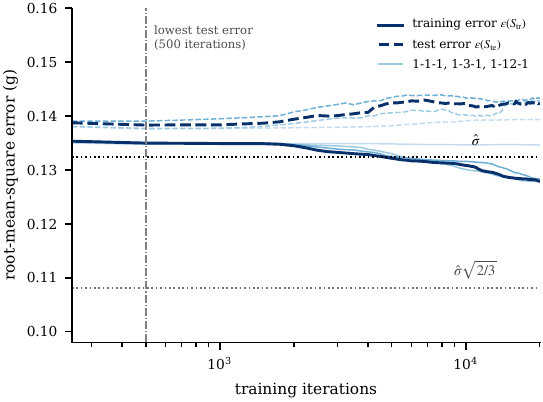}
 \caption{\textbf{Overfitting check.} Training error (solid) and test error (dashed),
 Eq.~\eqref{eq:rmse}, during training: root mean square over the twelve splits of scheme (b),
 with \NrandTest\ of the \Nweigh\ weighings held out. Dark: 1-6-1 networks; light: 1-1-1, 1-3-1 and
 1-12-1. Dotted lines: $\hat\sigma$, the smallest test error a model can reach, and
 $\hat\sigma\sqrt{2/3}$, the root-mean-square deviation of the weighings from their own mean.
 Dash-dotted line: iteration of the lowest 1-6-1 test error.}
 \label{fig:tt}
\end{figure}

A flexible model can keep lowering its training error by adapting to the noise of the particular
weighings it sees; its test error then grows. This is \emph{overfitting}, and it is detected by
recording both errors during training \cite{Roberts2017}. Figure~\ref{fig:tt} does so for the
twelve splits of scheme (b), and answers three questions.

\emph{Is the error of an acceptable size?} A model equal to the exact mean has test error
$\hat\sigma$, the scatter of a single weighing. The training weighings, instead, enter the means
with which they are compared: their root-mean-square deviation from their own mean is
$\hat\sigma\sqrt{2/3}$, so that a training error below $\hat\sigma$ is not a sign of overfitting.
The 1-6-1 networks end with a training error of \trainEnd~g, between the two levels, and a test
error of \teRanNNsix~g, \relSig\% above $\hat\sigma$.

\emph{Does longer training harm prediction?} The test error of the 1-6-1 networks is lowest after
\lcMinIt\ iterations; at 20\,000 iterations it is higher by \riseMean~mg on average (\riseRel\%),
with a standard deviation of \riseSD~mg between the splits. On twelve further splits, drawn
independently, the increase is \riseIndep~mg. Training to convergence therefore costs at most a few
milligrams on unseen weighings.

\emph{Why train to convergence?} Because the decomposition selected by the penalty is reached only
there. After \lcMinIt\ iterations the 1-6-1 networks trained on all the data return an uptake with
exponent $n_1=\earlyNone$ and a loss of \earlyL~g that reaches 1\% at \earlyTon~min; after
20\,000 iterations, $n_1=\convNone$ and a loss of \convL~g that reaches 1\% at \convTon~min. The
uptake is learned early and does not change. The smallest loss, which the penalty selects among
the decompositions of Eq.~\eqref{eq:inv}, is reached only as $\mathcal J$ converges; over the last
5000 iterations the training error changes by \trainConv\%.

\FloatBarrier

%% file: main.bbl
\begin{thebibliography}{55}%
\makeatletter
\providecommand \@ifxundefined [1]{%
 \@ifx{#1\undefined}
}%
\providecommand \@ifnum [1]{%
 \ifnum #1\expandafter \@firstoftwo
 \else \expandafter \@secondoftwo
 \fi
}%
\providecommand \@ifx [1]{%
 \ifx #1\expandafter \@firstoftwo
 \else \expandafter \@secondoftwo
 \fi
}%
\providecommand \natexlab [1]{#1}%
\providecommand \enquote  [1]{``#1''}%
\providecommand \bibnamefont  [1]{#1}%
\providecommand \bibfnamefont [1]{#1}%
\providecommand \citenamefont [1]{#1}%
\providecommand \href@noop [0]{\@secondoftwo}%
\providecommand \href [0]{\begingroup \@sanitize@url \@href}%
\providecommand \@href[1]{\@@startlink{#1}\@@href}%
\providecommand \@@href[1]{\endgroup#1\@@endlink}%
\providecommand \@sanitize@url [0]{\catcode `\\12\catcode `\$12\catcode
  `\&12\catcode `\#12\catcode `\^12\catcode `\_12\catcode `\%12\relax}%
\providecommand \@@startlink[1]{}%
\providecommand \@@endlink[0]{}%
\providecommand \url  [0]{\begingroup\@sanitize@url \@url }%
\providecommand \@url [1]{\endgroup\@href {#1}{\urlprefix }}%
\providecommand \urlprefix  [0]{URL }%
\providecommand \Eprint [0]{\href }%
\providecommand \doibase [0]{https://doi.org/}%
\providecommand \selectlanguage [0]{\@gobble}%
\providecommand \bibinfo  [0]{\@secondoftwo}%
\providecommand \bibfield  [0]{\@secondoftwo}%
\providecommand \translation [1]{[#1]}%
\providecommand \BibitemOpen [0]{}%
\providecommand \bibitemStop [0]{}%
\providecommand \bibitemNoStop [0]{.\EOS\space}%
\providecommand \EOS [0]{\spacefactor3000\relax}%
\providecommand \BibitemShut  [1]{\csname bibitem#1\endcsname}%
\let\auto@bib@innerbib\@empty
\bibitem [{\citenamefont {Simon}(1962)}]{Simon1962}%
  \BibitemOpen
  \bibfield  {author} {\bibinfo {author} {\bibfnamefont {H.~A.}\ \bibnamefont
  {Simon}},\ }\bibfield  {title} {\bibinfo {title} {The architecture of
  complexity},\ }\href@noop {} {\bibfield  {journal} {\bibinfo  {journal}
  {Proceedings of the American Philosophical Society}\ }\textbf {\bibinfo
  {volume} {106}},\ \bibinfo {pages} {467} (\bibinfo {year}
  {1962})}\BibitemShut {NoStop}%
\bibitem [{\citenamefont {Anderson}(1972)}]{Anderson1972}%
  \BibitemOpen
  \bibfield  {author} {\bibinfo {author} {\bibfnamefont {P.~W.}\ \bibnamefont
  {Anderson}},\ }\bibfield  {title} {\bibinfo {title} {More is different:
  Broken symmetry and the nature of the hierarchical structure of science},\
  }\href {https://doi.org/10.1126/science.177.4047.393} {\bibfield  {journal}
  {\bibinfo  {journal} {Science}\ }\textbf {\bibinfo {volume} {177}},\ \bibinfo
  {pages} {393} (\bibinfo {year} {1972})}\BibitemShut {NoStop}%
\bibitem [{\citenamefont {Ladyman}\ \emph {et~al.}(2013)\citenamefont
  {Ladyman}, \citenamefont {Lambert},\ and\ \citenamefont
  {Wiesner}}]{Ladyman2013}%
  \BibitemOpen
  \bibfield  {author} {\bibinfo {author} {\bibfnamefont {J.}~\bibnamefont
  {Ladyman}}, \bibinfo {author} {\bibfnamefont {J.}~\bibnamefont {Lambert}},\
  and\ \bibinfo {author} {\bibfnamefont {K.}~\bibnamefont {Wiesner}},\
  }\bibfield  {title} {\bibinfo {title} {What is a complex system?},\ }\href
  {https://doi.org/10.1007/s13194-012-0056-8} {\bibfield  {journal} {\bibinfo
  {journal} {European Journal for Philosophy of Science}\ }\textbf {\bibinfo
  {volume} {3}},\ \bibinfo {pages} {33} (\bibinfo {year} {2013})}\BibitemShut
  {NoStop}%
\bibitem [{\citenamefont {Ottino}(2004)}]{Ottino2004}%
  \BibitemOpen
  \bibfield  {author} {\bibinfo {author} {\bibfnamefont {J.~M.}\ \bibnamefont
  {Ottino}},\ }\bibfield  {title} {\bibinfo {title} {Engineering complex
  systems},\ }\href {https://doi.org/10.1038/427399a} {\bibfield  {journal}
  {\bibinfo  {journal} {Nature}\ }\textbf {\bibinfo {volume} {427}},\ \bibinfo
  {pages} {399} (\bibinfo {year} {2004})}\BibitemShut {NoStop}%
\bibitem [{\citenamefont {Gutenkunst}\ \emph {et~al.}(2007)\citenamefont
  {Gutenkunst}, \citenamefont {Waterfall}, \citenamefont {Casey}, \citenamefont
  {Brown}, \citenamefont {Myers},\ and\ \citenamefont
  {Sethna}}]{Gutenkunst2007}%
  \BibitemOpen
  \bibfield  {author} {\bibinfo {author} {\bibfnamefont {R.~N.}\ \bibnamefont
  {Gutenkunst}}, \bibinfo {author} {\bibfnamefont {J.~J.}\ \bibnamefont
  {Waterfall}}, \bibinfo {author} {\bibfnamefont {F.~P.}\ \bibnamefont
  {Casey}}, \bibinfo {author} {\bibfnamefont {K.~S.}\ \bibnamefont {Brown}},
  \bibinfo {author} {\bibfnamefont {C.~R.}\ \bibnamefont {Myers}},\ and\
  \bibinfo {author} {\bibfnamefont {J.~P.}\ \bibnamefont {Sethna}},\ }\bibfield
   {title} {\bibinfo {title} {Universally sloppy parameter sensitivities in
  systems biology models},\ }\href
  {https://doi.org/10.1371/journal.pcbi.0030189} {\bibfield  {journal}
  {\bibinfo  {journal} {PLoS Computational Biology}\ }\textbf {\bibinfo
  {volume} {3}},\ \bibinfo {pages} {e189} (\bibinfo {year} {2007})}\BibitemShut
  {NoStop}%
\bibitem [{\citenamefont {Raue}\ \emph {et~al.}(2009)\citenamefont {Raue},
  \citenamefont {Kreutz}, \citenamefont {Maiwald}, \citenamefont {Bachmann},
  \citenamefont {Schilling}, \citenamefont {Klingm{\"u}ller},\ and\
  \citenamefont {Timmer}}]{Raue2009}%
  \BibitemOpen
  \bibfield  {author} {\bibinfo {author} {\bibfnamefont {A.}~\bibnamefont
  {Raue}}, \bibinfo {author} {\bibfnamefont {C.}~\bibnamefont {Kreutz}},
  \bibinfo {author} {\bibfnamefont {T.}~\bibnamefont {Maiwald}}, \bibinfo
  {author} {\bibfnamefont {J.}~\bibnamefont {Bachmann}}, \bibinfo {author}
  {\bibfnamefont {M.}~\bibnamefont {Schilling}}, \bibinfo {author}
  {\bibfnamefont {U.}~\bibnamefont {Klingm{\"u}ller}},\ and\ \bibinfo {author}
  {\bibfnamefont {J.}~\bibnamefont {Timmer}},\ }\bibfield  {title} {\bibinfo
  {title} {Structural and practical identifiability analysis of partially
  observed dynamical models by exploiting the profile likelihood},\ }\href
  {https://doi.org/10.1093/bioinformatics/btp358} {\bibfield  {journal}
  {\bibinfo  {journal} {Bioinformatics}\ }\textbf {\bibinfo {volume} {25}},\
  \bibinfo {pages} {1923} (\bibinfo {year} {2009})}\BibitemShut {NoStop}%
\bibitem [{\citenamefont {Miao}\ \emph {et~al.}(2011)\citenamefont {Miao},
  \citenamefont {Xia}, \citenamefont {Perelson},\ and\ \citenamefont
  {Wu}}]{Miao2011}%
  \BibitemOpen
  \bibfield  {author} {\bibinfo {author} {\bibfnamefont {H.}~\bibnamefont
  {Miao}}, \bibinfo {author} {\bibfnamefont {X.}~\bibnamefont {Xia}}, \bibinfo
  {author} {\bibfnamefont {A.~S.}\ \bibnamefont {Perelson}},\ and\ \bibinfo
  {author} {\bibfnamefont {H.}~\bibnamefont {Wu}},\ }\bibfield  {title}
  {\bibinfo {title} {On identifiability of nonlinear {ODE} models and
  applications in viral dynamics},\ }\href {https://doi.org/10.1137/090757009}
  {\bibfield  {journal} {\bibinfo  {journal} {SIAM Review}\ }\textbf {\bibinfo
  {volume} {53}},\ \bibinfo {pages} {3} (\bibinfo {year} {2011})}\BibitemShut
  {NoStop}%
\bibitem [{\citenamefont {Transtrum}\ \emph {et~al.}(2015)\citenamefont
  {Transtrum}, \citenamefont {Machta}, \citenamefont {Brown}, \citenamefont
  {Daniels}, \citenamefont {Myers},\ and\ \citenamefont
  {Sethna}}]{Transtrum2015}%
  \BibitemOpen
  \bibfield  {author} {\bibinfo {author} {\bibfnamefont {M.~K.}\ \bibnamefont
  {Transtrum}}, \bibinfo {author} {\bibfnamefont {B.~B.}\ \bibnamefont
  {Machta}}, \bibinfo {author} {\bibfnamefont {K.~S.}\ \bibnamefont {Brown}},
  \bibinfo {author} {\bibfnamefont {B.~C.}\ \bibnamefont {Daniels}}, \bibinfo
  {author} {\bibfnamefont {C.~R.}\ \bibnamefont {Myers}},\ and\ \bibinfo
  {author} {\bibfnamefont {J.~P.}\ \bibnamefont {Sethna}},\ }\bibfield  {title}
  {\bibinfo {title} {Perspective: Sloppiness and emergent theories in physics,
  biology, and beyond},\ }\href {https://doi.org/10.1063/1.4923066} {\bibfield
  {journal} {\bibinfo  {journal} {The Journal of Chemical Physics}\ }\textbf
  {\bibinfo {volume} {143}},\ \bibinfo {pages} {010901} (\bibinfo {year}
  {2015})}\BibitemShut {NoStop}%
\bibitem [{\citenamefont {Leopold}(1983)}]{Leopold1983}%
  \BibitemOpen
  \bibfield  {author} {\bibinfo {author} {\bibfnamefont {A.~C.}\ \bibnamefont
  {Leopold}},\ }\bibfield  {title} {\bibinfo {title} {Volumetric components of
  seed imbibition},\ }\href {https://doi.org/10.1104/pp.73.3.677} {\bibfield
  {journal} {\bibinfo  {journal} {Plant Physiology}\ }\textbf {\bibinfo
  {volume} {73}},\ \bibinfo {pages} {677} (\bibinfo {year} {1983})}\BibitemShut
  {NoStop}%
\bibitem [{\citenamefont {Louf}\ \emph {et~al.}(2018)\citenamefont {Louf},
  \citenamefont {Zheng}, \citenamefont {Kumar}, \citenamefont {Bohr},
  \citenamefont {Gundlach}, \citenamefont {Harholt}, \citenamefont {Poulsen},\
  and\ \citenamefont {Jensen}}]{Louf2018}%
  \BibitemOpen
  \bibfield  {author} {\bibinfo {author} {\bibfnamefont {J.-F.}\ \bibnamefont
  {Louf}}, \bibinfo {author} {\bibfnamefont {Y.}~\bibnamefont {Zheng}},
  \bibinfo {author} {\bibfnamefont {A.}~\bibnamefont {Kumar}}, \bibinfo
  {author} {\bibfnamefont {T.}~\bibnamefont {Bohr}}, \bibinfo {author}
  {\bibfnamefont {C.}~\bibnamefont {Gundlach}}, \bibinfo {author}
  {\bibfnamefont {J.}~\bibnamefont {Harholt}}, \bibinfo {author} {\bibfnamefont
  {H.~F.}\ \bibnamefont {Poulsen}},\ and\ \bibinfo {author} {\bibfnamefont
  {K.~H.}\ \bibnamefont {Jensen}},\ }\bibfield  {title} {\bibinfo {title}
  {Imbibition in plant seeds},\ }\href
  {https://doi.org/10.1103/PhysRevE.98.042403} {\bibfield  {journal} {\bibinfo
  {journal} {Physical Review E}\ }\textbf {\bibinfo {volume} {98}},\ \bibinfo
  {pages} {042403} (\bibinfo {year} {2018})}\BibitemShut {NoStop}%
\bibitem [{\citenamefont {Munz}\ \emph {et~al.}(2017)\citenamefont {Munz},
  \citenamefont {Rolletschek}, \citenamefont {Oeltze-Jafra}, \citenamefont
  {Fuchs}, \citenamefont {Guendel}, \citenamefont {Neuberger}, \citenamefont
  {Ortleb}, \citenamefont {Jakob},\ and\ \citenamefont {Borisjuk}}]{Munz2017}%
  \BibitemOpen
  \bibfield  {author} {\bibinfo {author} {\bibfnamefont {E.}~\bibnamefont
  {Munz}}, \bibinfo {author} {\bibfnamefont {H.}~\bibnamefont {Rolletschek}},
  \bibinfo {author} {\bibfnamefont {S.}~\bibnamefont {Oeltze-Jafra}}, \bibinfo
  {author} {\bibfnamefont {J.}~\bibnamefont {Fuchs}}, \bibinfo {author}
  {\bibfnamefont {A.}~\bibnamefont {Guendel}}, \bibinfo {author} {\bibfnamefont
  {T.}~\bibnamefont {Neuberger}}, \bibinfo {author} {\bibfnamefont
  {S.}~\bibnamefont {Ortleb}}, \bibinfo {author} {\bibfnamefont {P.~M.}\
  \bibnamefont {Jakob}},\ and\ \bibinfo {author} {\bibfnamefont
  {L.}~\bibnamefont {Borisjuk}},\ }\bibfield  {title} {\bibinfo {title} {A
  functional imaging study of germinating oilseed rape seed},\ }\href
  {https://doi.org/10.1111/nph.14736} {\bibfield  {journal} {\bibinfo
  {journal} {New Phytologist}\ }\textbf {\bibinfo {volume} {216}},\ \bibinfo
  {pages} {1181} (\bibinfo {year} {2017})}\BibitemShut {NoStop}%
\bibitem [{\citenamefont {Lechowska}\ \emph {et~al.}(2019)\citenamefont
  {Lechowska}, \citenamefont {Kubala}, \citenamefont {Wojtyla}, \citenamefont
  {Nowaczyk}, \citenamefont {Quinet}, \citenamefont {Lutts},\ and\
  \citenamefont {Garnczarska}}]{Lechowska2019}%
  \BibitemOpen
  \bibfield  {author} {\bibinfo {author} {\bibfnamefont {K.}~\bibnamefont
  {Lechowska}}, \bibinfo {author} {\bibfnamefont {S.}~\bibnamefont {Kubala}},
  \bibinfo {author} {\bibfnamefont {{\L}.}~\bibnamefont {Wojtyla}}, \bibinfo
  {author} {\bibfnamefont {G.}~\bibnamefont {Nowaczyk}}, \bibinfo {author}
  {\bibfnamefont {M.}~\bibnamefont {Quinet}}, \bibinfo {author} {\bibfnamefont
  {S.}~\bibnamefont {Lutts}},\ and\ \bibinfo {author} {\bibfnamefont
  {M.}~\bibnamefont {Garnczarska}},\ }\bibfield  {title} {\bibinfo {title} {New
  insight on water status in germinating {Brassica napus} seeds in relation to
  priming-improved germination},\ }\href {https://doi.org/10.3390/ijms20030540}
  {\bibfield  {journal} {\bibinfo  {journal} {International Journal of
  Molecular Sciences}\ }\textbf {\bibinfo {volume} {20}},\ \bibinfo {pages}
  {540} (\bibinfo {year} {2019})}\BibitemShut {NoStop}%
\bibitem [{\citenamefont {Aangenendt}\ \emph {et~al.}(2020)\citenamefont
  {Aangenendt}, \citenamefont {Punter}, \citenamefont {Mulder},\ and\
  \citenamefont {Wyss}}]{Aangenendt2020}%
  \BibitemOpen
  \bibfield  {author} {\bibinfo {author} {\bibfnamefont {F.~J.}\ \bibnamefont
  {Aangenendt}}, \bibinfo {author} {\bibfnamefont {M.~T. J. J.~M.}\
  \bibnamefont {Punter}}, \bibinfo {author} {\bibfnamefont {B.~M.}\
  \bibnamefont {Mulder}},\ and\ \bibinfo {author} {\bibfnamefont {H.~M.}\
  \bibnamefont {Wyss}},\ }\bibfield  {title} {\bibinfo {title} {Nonmonotonic
  swelling and compression dynamics of hydrogels in polymer solutions},\ }\href
  {https://doi.org/10.1103/PhysRevE.102.062606} {\bibfield  {journal} {\bibinfo
   {journal} {Physical Review E}\ }\textbf {\bibinfo {volume} {102}},\ \bibinfo
  {pages} {062606} (\bibinfo {year} {2020})}\BibitemShut {NoStop}%
\bibitem [{\citenamefont {Chen}\ and\ \citenamefont
  {Ravi-Chandar}(2022)}]{ChenRaviChandar2022}%
  \BibitemOpen
  \bibfield  {author} {\bibinfo {author} {\bibfnamefont {S.}~\bibnamefont
  {Chen}}\ and\ \bibinfo {author} {\bibfnamefont {K.}~\bibnamefont
  {Ravi-Chandar}},\ }\bibfield  {title} {\bibinfo {title} {Nonlinear
  poroviscoelastic behavior of gelatin-based hydrogel},\ }\href
  {https://doi.org/10.1016/j.jmps.2021.104650} {\bibfield  {journal} {\bibinfo
  {journal} {Journal of the Mechanics and Physics of Solids}\ }\textbf
  {\bibinfo {volume} {158}},\ \bibinfo {pages} {104650} (\bibinfo {year}
  {2022})}\BibitemShut {NoStop}%
\bibitem [{\citenamefont {Machado}\ \emph {et~al.}(2026)\citenamefont
  {Machado}, \citenamefont {Lima-Ten{\'o}rio},\ and\ \citenamefont
  {Ten{\'o}rio-Neto}}]{Machado2026}%
  \BibitemOpen
  \bibfield  {author} {\bibinfo {author} {\bibfnamefont {V.~D.}\ \bibnamefont
  {Machado}}, \bibinfo {author} {\bibfnamefont {M.~K.}\ \bibnamefont
  {Lima-Ten{\'o}rio}},\ and\ \bibinfo {author} {\bibfnamefont {E.~T.}\
  \bibnamefont {Ten{\'o}rio-Neto}},\ }\bibfield  {title} {\bibinfo {title}
  {Understanding the swelling behavior of polysaccharide-based hydrogels
  through a kinetic modeling},\ }\href
  {https://doi.org/10.1021/acspolymersau.5c00208} {\bibfield  {journal}
  {\bibinfo  {journal} {ACS Polymers Au}\ }\textbf {\bibinfo {volume} {6}},\
  \bibinfo {pages} {634} (\bibinfo {year} {2026})}\BibitemShut {NoStop}%
\bibitem [{\citenamefont {Michailova}\ \emph {et~al.}(2000)\citenamefont
  {Michailova}, \citenamefont {Titeva}, \citenamefont {Kotsilkova},
  \citenamefont {Krusteva},\ and\ \citenamefont {Minkov}}]{Michailova2000}%
  \BibitemOpen
  \bibfield  {author} {\bibinfo {author} {\bibfnamefont {V.}~\bibnamefont
  {Michailova}}, \bibinfo {author} {\bibfnamefont {S.}~\bibnamefont {Titeva}},
  \bibinfo {author} {\bibfnamefont {R.}~\bibnamefont {Kotsilkova}}, \bibinfo
  {author} {\bibfnamefont {E.}~\bibnamefont {Krusteva}},\ and\ \bibinfo
  {author} {\bibfnamefont {E.}~\bibnamefont {Minkov}},\ }\bibfield  {title}
  {\bibinfo {title} {Water uptake and relaxation processes in mixed unlimited
  swelling hydrogels},\ }\href {https://doi.org/10.1016/S0378-5173(00)00536-6}
  {\bibfield  {journal} {\bibinfo  {journal} {International Journal of
  Pharmaceutics}\ }\textbf {\bibinfo {volume} {209}},\ \bibinfo {pages} {45}
  (\bibinfo {year} {2000})}\BibitemShut {NoStop}%
\bibitem [{\citenamefont {Peleg}(1988)}]{Peleg1988}%
  \BibitemOpen
  \bibfield  {author} {\bibinfo {author} {\bibfnamefont {M.}~\bibnamefont
  {Peleg}},\ }\bibfield  {title} {\bibinfo {title} {An empirical model for the
  description of moisture sorption curves},\ }\href
  {https://doi.org/10.1111/j.1365-2621.1988.tb13565.x} {\bibfield  {journal}
  {\bibinfo  {journal} {Journal of Food Science}\ }\textbf {\bibinfo {volume}
  {53}},\ \bibinfo {pages} {1216} (\bibinfo {year} {1988})}\BibitemShut
  {NoStop}%
\bibitem [{\citenamefont {Weibull}(1951)}]{Weibull1951}%
  \BibitemOpen
  \bibfield  {author} {\bibinfo {author} {\bibfnamefont {W.}~\bibnamefont
  {Weibull}},\ }\bibfield  {title} {\bibinfo {title} {A statistical
  distribution function of wide applicability},\ }\href
  {https://doi.org/10.1115/1.4010337} {\bibfield  {journal} {\bibinfo
  {journal} {Journal of Applied Mechanics}\ }\textbf {\bibinfo {volume} {18}},\
  \bibinfo {pages} {293} (\bibinfo {year} {1951})}\BibitemShut {NoStop}%
\bibitem [{\citenamefont {Machado}\ \emph {et~al.}(1999)\citenamefont
  {Machado}, \citenamefont {Oliveira},\ and\ \citenamefont
  {Cunha}}]{MachadoOliveira1999}%
  \BibitemOpen
  \bibfield  {author} {\bibinfo {author} {\bibfnamefont {M.~F.}\ \bibnamefont
  {Machado}}, \bibinfo {author} {\bibfnamefont {F.~A.~R.}\ \bibnamefont
  {Oliveira}},\ and\ \bibinfo {author} {\bibfnamefont {L.~M.}\ \bibnamefont
  {Cunha}},\ }\bibfield  {title} {\bibinfo {title} {Effect of milk fat and
  total solids concentration on the kinetics of moisture uptake by ready-to-eat
  breakfast cereal},\ }\href {https://doi.org/10.1046/j.1365-2621.1999.00238.x}
  {\bibfield  {journal} {\bibinfo  {journal} {International Journal of Food
  Science and Technology}\ }\textbf {\bibinfo {volume} {34}},\ \bibinfo {pages}
  {47} (\bibinfo {year} {1999})}\BibitemShut {NoStop}%
\bibitem [{\citenamefont {Marabi}\ \emph {et~al.}(2003)\citenamefont {Marabi},
  \citenamefont {Livings}, \citenamefont {Jacobson},\ and\ \citenamefont
  {Saguy}}]{Marabi2003}%
  \BibitemOpen
  \bibfield  {author} {\bibinfo {author} {\bibfnamefont {A.}~\bibnamefont
  {Marabi}}, \bibinfo {author} {\bibfnamefont {S.}~\bibnamefont {Livings}},
  \bibinfo {author} {\bibfnamefont {M.}~\bibnamefont {Jacobson}},\ and\
  \bibinfo {author} {\bibfnamefont {I.~S.}\ \bibnamefont {Saguy}},\ }\bibfield
  {title} {\bibinfo {title} {Normalized {W}eibull distribution for modeling
  rehydration of food particulates},\ }\href
  {https://doi.org/10.1007/s00217-003-0719-y} {\bibfield  {journal} {\bibinfo
  {journal} {European Food Research and Technology}\ }\textbf {\bibinfo
  {volume} {217}},\ \bibinfo {pages} {311} (\bibinfo {year}
  {2003})}\BibitemShut {NoStop}%
\bibitem [{\citenamefont {Rafiq}\ \emph {et~al.}(2015)\citenamefont {Rafiq},
  \citenamefont {Chowdhary}, \citenamefont {Hazarika},\ and\ \citenamefont
  {Makroo}}]{Rafiq2015}%
  \BibitemOpen
  \bibfield  {author} {\bibinfo {author} {\bibfnamefont {A.}~\bibnamefont
  {Rafiq}}, \bibinfo {author} {\bibfnamefont {J.}~\bibnamefont {Chowdhary}},
  \bibinfo {author} {\bibfnamefont {M.~K.}\ \bibnamefont {Hazarika}},\ and\
  \bibinfo {author} {\bibfnamefont {H.~A.}\ \bibnamefont {Makroo}},\ }\bibfield
   {title} {\bibinfo {title} {Temperature dependence on hydration kinetic model
  parameters during rehydration of parboiled rice},\ }\href
  {https://doi.org/10.1007/s13197-015-1790-7} {\bibfield  {journal} {\bibinfo
  {journal} {Journal of Food Science and Technology}\ }\textbf {\bibinfo
  {volume} {52}},\ \bibinfo {pages} {6090} (\bibinfo {year}
  {2015})}\BibitemShut {NoStop}%
\bibitem [{\citenamefont {Crank}(1975)}]{Crank1975}%
  \BibitemOpen
  \bibfield  {author} {\bibinfo {author} {\bibfnamefont {J.}~\bibnamefont
  {Crank}},\ }\href@noop {} {\emph {\bibinfo {title} {The Mathematics of
  Diffusion}}},\ \bibinfo {edition} {2nd}\ ed.\ (\bibinfo  {publisher}
  {Clarendon Press},\ \bibinfo {address} {Oxford},\ \bibinfo {year}
  {1975})\BibitemShut {NoStop}%
\bibitem [{\citenamefont {Hsu}(1983)}]{Hsu1983}%
  \BibitemOpen
  \bibfield  {author} {\bibinfo {author} {\bibfnamefont {K.~H.}\ \bibnamefont
  {Hsu}},\ }\bibfield  {title} {\bibinfo {title} {A diffusion model with a
  concentration-dependent diffusion coefficient for describing water movement
  in legumes during soaking},\ }\href@noop {} {\bibfield  {journal} {\bibinfo
  {journal} {Journal of Food Science}\ }\textbf {\bibinfo {volume} {48}},\
  \bibinfo {pages} {618} (\bibinfo {year} {1983})}\BibitemShut {NoStop}%
\bibitem [{\citenamefont {Peppas}\ and\ \citenamefont
  {Brannon-Peppas}(1994)}]{Peppas1994}%
  \BibitemOpen
  \bibfield  {author} {\bibinfo {author} {\bibfnamefont {N.~A.}\ \bibnamefont
  {Peppas}}\ and\ \bibinfo {author} {\bibfnamefont {L.}~\bibnamefont
  {Brannon-Peppas}},\ }\bibfield  {title} {\bibinfo {title} {Water diffusion
  and sorption in amorphous macromolecular systems and foods},\ }\href
  {https://doi.org/10.1016/0260-8774(94)90030-2} {\bibfield  {journal}
  {\bibinfo  {journal} {Journal of Food Engineering}\ }\textbf {\bibinfo
  {volume} {22}},\ \bibinfo {pages} {189} (\bibinfo {year} {1994})}\BibitemShut
  {NoStop}%
\bibitem [{\citenamefont {Rubinstein}(1971)}]{rubinshteuin1971stefan}%
  \BibitemOpen
  \bibfield  {author} {\bibinfo {author} {\bibfnamefont {L.~I.}\ \bibnamefont
  {Rubinstein}},\ }\href@noop {} {\emph {\bibinfo {title} {The {S}tefan
  Problem}}},\ \bibinfo {series} {Translations of Mathematical Monographs},
  Vol.~\bibinfo {volume} {27}\ (\bibinfo  {publisher} {American Mathematical
  Society},\ \bibinfo {address} {Providence, RI},\ \bibinfo {year}
  {1971})\BibitemShut {NoStop}%
\bibitem [{\citenamefont {Fasano}\ and\ \citenamefont
  {Primicerio}(1983)}]{fasano1983free}%
  \BibitemOpen
  \bibinfo {editor} {\bibfnamefont {A.}~\bibnamefont {Fasano}}\ and\ \bibinfo
  {editor} {\bibfnamefont {M.}~\bibnamefont {Primicerio}},\ eds.,\ \href@noop
  {} {\emph {\bibinfo {title} {Free Boundary Problems: Theory and
  Applications}}},\ \bibinfo {series} {Research Notes in Mathematics}, Vol.\
  \bibinfo {volume} {78--79}\ (\bibinfo  {publisher} {Pitman},\ \bibinfo
  {address} {Boston},\ \bibinfo {year} {1983})\BibitemShut {NoStop}%
\bibitem [{\citenamefont {Me{\u\i}rmanov}(1992)}]{meuirmanov1992stefan}%
  \BibitemOpen
  \bibfield  {author} {\bibinfo {author} {\bibfnamefont {A.~M.}\ \bibnamefont
  {Me{\u\i}rmanov}},\ }\href@noop {} {\emph {\bibinfo {title} {The {S}tefan
  Problem}}},\ \bibinfo {series} {De Gruyter Expositions in Mathematics},
  Vol.~\bibinfo {volume} {3}\ (\bibinfo  {publisher} {Walter de Gruyter},\
  \bibinfo {year} {1992})\BibitemShut {NoStop}%
\bibitem [{\citenamefont {Meacci}\ \emph {et~al.}(2026)\citenamefont {Meacci},
  \citenamefont {Farina}, \citenamefont {Fusi}, \citenamefont {Gianni},
  \citenamefont {{di Bari}},\ and\ \citenamefont
  {Aliyari}}]{Meacci2026FreeBoundary}%
  \BibitemOpen
  \bibfield  {author} {\bibinfo {author} {\bibfnamefont {L.}~\bibnamefont
  {Meacci}}, \bibinfo {author} {\bibfnamefont {A.}~\bibnamefont {Farina}},
  \bibinfo {author} {\bibfnamefont {L.}~\bibnamefont {Fusi}}, \bibinfo {author}
  {\bibfnamefont {R.}~\bibnamefont {Gianni}}, \bibinfo {author} {\bibfnamefont
  {V.}~\bibnamefont {{di Bari}}},\ and\ \bibinfo {author} {\bibfnamefont
  {A.}~\bibnamefont {Aliyari}},\ }\bibfield  {title} {\bibinfo {title}
  {Imbibition in plant seeds: A free-boundary universal framework for swelling
  porous media}} (\bibinfo {year} {2026}),\ \bibinfo {note} {manuscript under
  review at Physical Review Letters}\BibitemShut {NoStop}%
\bibitem [{\citenamefont {Atkinson}\ and\ \citenamefont
  {Fedorov}(1975)}]{AtkinsonFedorov1975}%
  \BibitemOpen
  \bibfield  {author} {\bibinfo {author} {\bibfnamefont {A.~C.}\ \bibnamefont
  {Atkinson}}\ and\ \bibinfo {author} {\bibfnamefont {V.~V.}\ \bibnamefont
  {Fedorov}},\ }\bibfield  {title} {\bibinfo {title} {The design of experiments
  for discriminating between two rival models},\ }\href
  {https://doi.org/10.1093/biomet/62.1.57} {\bibfield  {journal} {\bibinfo
  {journal} {Biometrika}\ }\textbf {\bibinfo {volume} {62}},\ \bibinfo {pages}
  {57} (\bibinfo {year} {1975})}\BibitemShut {NoStop}%
\bibitem [{\citenamefont {Apgar}\ \emph {et~al.}(2008)\citenamefont {Apgar},
  \citenamefont {Toettcher}, \citenamefont {Endy}, \citenamefont {White},\ and\
  \citenamefont {Tidor}}]{Apgar2008}%
  \BibitemOpen
  \bibfield  {author} {\bibinfo {author} {\bibfnamefont {J.~F.}\ \bibnamefont
  {Apgar}}, \bibinfo {author} {\bibfnamefont {J.~E.}\ \bibnamefont
  {Toettcher}}, \bibinfo {author} {\bibfnamefont {D.}~\bibnamefont {Endy}},
  \bibinfo {author} {\bibfnamefont {F.~M.}\ \bibnamefont {White}},\ and\
  \bibinfo {author} {\bibfnamefont {B.}~\bibnamefont {Tidor}},\ }\bibfield
  {title} {\bibinfo {title} {Stimulus design for model selection and validation
  in cell signaling},\ }\href {https://doi.org/10.1371/journal.pcbi.0040030}
  {\bibfield  {journal} {\bibinfo  {journal} {PLoS Computational Biology}\
  }\textbf {\bibinfo {volume} {4}},\ \bibinfo {pages} {e30} (\bibinfo {year}
  {2008})}\BibitemShut {NoStop}%
\bibitem [{\citenamefont {Skanda}\ and\ \citenamefont
  {Lebiedz}(2010)}]{SkandaLebiedz2010}%
  \BibitemOpen
  \bibfield  {author} {\bibinfo {author} {\bibfnamefont {D.}~\bibnamefont
  {Skanda}}\ and\ \bibinfo {author} {\bibfnamefont {D.}~\bibnamefont
  {Lebiedz}},\ }\bibfield  {title} {\bibinfo {title} {An optimal experimental
  design approach to model discrimination in dynamic biochemical systems},\
  }\href {https://doi.org/10.1093/bioinformatics/btq074} {\bibfield  {journal}
  {\bibinfo  {journal} {Bioinformatics}\ }\textbf {\bibinfo {volume} {26}},\
  \bibinfo {pages} {939} (\bibinfo {year} {2010})}\BibitemShut {NoStop}%
\bibitem [{\citenamefont {Lorenz}\ \emph {et~al.}(2007)\citenamefont {Lorenz},
  \citenamefont {Diederichs}, \citenamefont {Telgmann},\ and\ \citenamefont
  {Sch{\"u}tte}}]{Lorenz2007}%
  \BibitemOpen
  \bibfield  {author} {\bibinfo {author} {\bibfnamefont {S.}~\bibnamefont
  {Lorenz}}, \bibinfo {author} {\bibfnamefont {E.}~\bibnamefont {Diederichs}},
  \bibinfo {author} {\bibfnamefont {R.}~\bibnamefont {Telgmann}},\ and\
  \bibinfo {author} {\bibfnamefont {C.}~\bibnamefont {Sch{\"u}tte}},\
  }\bibfield  {title} {\bibinfo {title} {Discrimination of dynamical system
  models for biological and chemical processes},\ }\href
  {https://doi.org/10.1002/jcc.20674} {\bibfield  {journal} {\bibinfo
  {journal} {Journal of Computational Chemistry}\ }\textbf {\bibinfo {volume}
  {28}},\ \bibinfo {pages} {1384} (\bibinfo {year} {2007})}\BibitemShut
  {NoStop}%
\bibitem [{\citenamefont {Hamadeh}\ \emph {et~al.}(2013)\citenamefont
  {Hamadeh}, \citenamefont {Ingalls},\ and\ \citenamefont
  {Sontag}}]{Hamadeh2013}%
  \BibitemOpen
  \bibfield  {author} {\bibinfo {author} {\bibfnamefont {A.}~\bibnamefont
  {Hamadeh}}, \bibinfo {author} {\bibfnamefont {B.}~\bibnamefont {Ingalls}},\
  and\ \bibinfo {author} {\bibfnamefont {E.~D.}\ \bibnamefont {Sontag}},\
  }\bibfield  {title} {\bibinfo {title} {Transient dynamic phenotypes as
  criteria for model discrimination: Fold-change detection in {Rhodobacter
  sphaeroides} chemotaxis},\ }\href {https://doi.org/10.1098/rsif.2012.0935}
  {\bibfield  {journal} {\bibinfo  {journal} {Journal of the Royal Society
  Interface}\ }\textbf {\bibinfo {volume} {10}},\ \bibinfo {pages} {20120935}
  (\bibinfo {year} {2013})}\BibitemShut {NoStop}%
\bibitem [{\citenamefont {Sontag}(2025)}]{Sontag2025}%
  \BibitemOpen
  \bibfield  {author} {\bibinfo {author} {\bibfnamefont {E.~D.}\ \bibnamefont
  {Sontag}},\ }\bibfield  {title} {\bibinfo {title} {Dynamic response
  phenotypes and model discrimination in systems and synthetic biology},\
  }\href@noop {} {\bibfield  {journal} {\bibinfo  {journal} {arXiv preprint
  arXiv:2512.24945}\ } (\bibinfo {year} {2025})}\BibitemShut {NoStop}%
\bibitem [{\citenamefont {Rackauckas}\ \emph {et~al.}(2020)\citenamefont
  {Rackauckas}, \citenamefont {Ma}, \citenamefont {Martensen}, \citenamefont
  {Warner}, \citenamefont {Zubov}, \citenamefont {Supekar}, \citenamefont
  {Skinner}, \citenamefont {Ramadhan},\ and\ \citenamefont
  {Edelman}}]{Rackauckas2020}%
  \BibitemOpen
  \bibfield  {author} {\bibinfo {author} {\bibfnamefont {C.}~\bibnamefont
  {Rackauckas}}, \bibinfo {author} {\bibfnamefont {Y.}~\bibnamefont {Ma}},
  \bibinfo {author} {\bibfnamefont {J.}~\bibnamefont {Martensen}}, \bibinfo
  {author} {\bibfnamefont {C.}~\bibnamefont {Warner}}, \bibinfo {author}
  {\bibfnamefont {K.}~\bibnamefont {Zubov}}, \bibinfo {author} {\bibfnamefont
  {R.}~\bibnamefont {Supekar}}, \bibinfo {author} {\bibfnamefont
  {D.}~\bibnamefont {Skinner}}, \bibinfo {author} {\bibfnamefont
  {A.}~\bibnamefont {Ramadhan}},\ and\ \bibinfo {author} {\bibfnamefont
  {A.}~\bibnamefont {Edelman}},\ }\bibfield  {title} {\bibinfo {title}
  {Universal differential equations for scientific machine learning},\
  }\href@noop {} {\bibfield  {journal} {\bibinfo  {journal} {arXiv preprint
  arXiv:2001.04385}\ } (\bibinfo {year} {2020})}\BibitemShut {NoStop}%
\bibitem [{\citenamefont {Karniadakis}\ \emph {et~al.}(2021)\citenamefont
  {Karniadakis}, \citenamefont {Kevrekidis}, \citenamefont {Lu}, \citenamefont
  {Perdikaris}, \citenamefont {Wang},\ and\ \citenamefont
  {Yang}}]{Karniadakis2021}%
  \BibitemOpen
  \bibfield  {author} {\bibinfo {author} {\bibfnamefont {G.~E.}\ \bibnamefont
  {Karniadakis}}, \bibinfo {author} {\bibfnamefont {I.~G.}\ \bibnamefont
  {Kevrekidis}}, \bibinfo {author} {\bibfnamefont {L.}~\bibnamefont {Lu}},
  \bibinfo {author} {\bibfnamefont {P.}~\bibnamefont {Perdikaris}}, \bibinfo
  {author} {\bibfnamefont {S.}~\bibnamefont {Wang}},\ and\ \bibinfo {author}
  {\bibfnamefont {L.}~\bibnamefont {Yang}},\ }\bibfield  {title} {\bibinfo
  {title} {Physics-informed machine learning},\ }\href
  {https://doi.org/10.1038/s42254-021-00314-5} {\bibfield  {journal} {\bibinfo
  {journal} {Nature Reviews Physics}\ }\textbf {\bibinfo {volume} {3}},\
  \bibinfo {pages} {422} (\bibinfo {year} {2021})}\BibitemShut {NoStop}%
\bibitem [{\citenamefont {Masi}\ \emph {et~al.}(2021)\citenamefont {Masi},
  \citenamefont {Stefanou}, \citenamefont {Vannucci},\ and\ \citenamefont
  {Maffi-Berthier}}]{Masi2021}%
  \BibitemOpen
  \bibfield  {author} {\bibinfo {author} {\bibfnamefont {F.}~\bibnamefont
  {Masi}}, \bibinfo {author} {\bibfnamefont {I.}~\bibnamefont {Stefanou}},
  \bibinfo {author} {\bibfnamefont {P.}~\bibnamefont {Vannucci}},\ and\
  \bibinfo {author} {\bibfnamefont {V.}~\bibnamefont {Maffi-Berthier}},\
  }\bibfield  {title} {\bibinfo {title} {Thermodynamics-based artificial neural
  networks for constitutive modeling},\ }\href
  {https://doi.org/10.1016/j.jmps.2020.104277} {\bibfield  {journal} {\bibinfo
  {journal} {Journal of the Mechanics and Physics of Solids}\ }\textbf
  {\bibinfo {volume} {147}},\ \bibinfo {pages} {104277} (\bibinfo {year}
  {2021})}\BibitemShut {NoStop}%
\bibitem [{\citenamefont {Flaschel}\ \emph {et~al.}(2021)\citenamefont
  {Flaschel}, \citenamefont {Kumar},\ and\ \citenamefont
  {De~Lorenzis}}]{Flaschel2021}%
  \BibitemOpen
  \bibfield  {author} {\bibinfo {author} {\bibfnamefont {M.}~\bibnamefont
  {Flaschel}}, \bibinfo {author} {\bibfnamefont {S.}~\bibnamefont {Kumar}},\
  and\ \bibinfo {author} {\bibfnamefont {L.}~\bibnamefont {De~Lorenzis}},\
  }\bibfield  {title} {\bibinfo {title} {Unsupervised discovery of
  interpretable hyperelastic constitutive laws},\ }\href
  {https://doi.org/10.1016/j.cma.2021.113852} {\bibfield  {journal} {\bibinfo
  {journal} {Computer Methods in Applied Mechanics and Engineering}\ }\textbf
  {\bibinfo {volume} {381}},\ \bibinfo {pages} {113852} (\bibinfo {year}
  {2021})}\BibitemShut {NoStop}%
\bibitem [{\citenamefont {Linka}\ and\ \citenamefont
  {Kuhl}(2023)}]{LinkaKuhl2023}%
  \BibitemOpen
  \bibfield  {author} {\bibinfo {author} {\bibfnamefont {K.}~\bibnamefont
  {Linka}}\ and\ \bibinfo {author} {\bibfnamefont {E.}~\bibnamefont {Kuhl}},\
  }\bibfield  {title} {\bibinfo {title} {A new family of constitutive
  artificial neural networks towards automated model discovery},\ }\href
  {https://doi.org/10.1016/j.cma.2022.115731} {\bibfield  {journal} {\bibinfo
  {journal} {Computer Methods in Applied Mechanics and Engineering}\ }\textbf
  {\bibinfo {volume} {403}},\ \bibinfo {pages} {115731} (\bibinfo {year}
  {2023})}\BibitemShut {NoStop}%
\bibitem [{\citenamefont {Brunton}\ \emph {et~al.}(2016)\citenamefont
  {Brunton}, \citenamefont {Proctor},\ and\ \citenamefont
  {Kutz}}]{Brunton2016}%
  \BibitemOpen
  \bibfield  {author} {\bibinfo {author} {\bibfnamefont {S.~L.}\ \bibnamefont
  {Brunton}}, \bibinfo {author} {\bibfnamefont {J.~L.}\ \bibnamefont
  {Proctor}},\ and\ \bibinfo {author} {\bibfnamefont {J.~N.}\ \bibnamefont
  {Kutz}},\ }\bibfield  {title} {\bibinfo {title} {Discovering governing
  equations from data by sparse identification of nonlinear dynamical
  systems},\ }\href {https://doi.org/10.1073/pnas.1517384113} {\bibfield
  {journal} {\bibinfo  {journal} {Proceedings of the National Academy of
  Sciences}\ }\textbf {\bibinfo {volume} {113}},\ \bibinfo {pages} {3932}
  (\bibinfo {year} {2016})}\BibitemShut {NoStop}%
\bibitem [{\citenamefont {Chen}\ \emph {et~al.}(2018)\citenamefont {Chen},
  \citenamefont {Rubanova}, \citenamefont {Bettencourt},\ and\ \citenamefont
  {Duvenaud}}]{Chen2018}%
  \BibitemOpen
  \bibfield  {author} {\bibinfo {author} {\bibfnamefont {R.~T.~Q.}\
  \bibnamefont {Chen}}, \bibinfo {author} {\bibfnamefont {Y.}~\bibnamefont
  {Rubanova}}, \bibinfo {author} {\bibfnamefont {J.}~\bibnamefont
  {Bettencourt}},\ and\ \bibinfo {author} {\bibfnamefont {D.~K.}\ \bibnamefont
  {Duvenaud}},\ }\bibfield  {title} {\bibinfo {title} {Neural ordinary
  differential equations},\ }\href@noop {} {\bibfield  {journal} {\bibinfo
  {journal} {Advances in Neural Information Processing Systems}\ }\textbf
  {\bibinfo {volume} {31}} (\bibinfo {year} {2018})}\BibitemShut {NoStop}%
\bibitem [{\citenamefont {Wang}\ and\ \citenamefont
  {Perdikaris}(2021)}]{WangPerdikaris2021}%
  \BibitemOpen
  \bibfield  {author} {\bibinfo {author} {\bibfnamefont {S.}~\bibnamefont
  {Wang}}\ and\ \bibinfo {author} {\bibfnamefont {P.}~\bibnamefont
  {Perdikaris}},\ }\bibfield  {title} {\bibinfo {title} {Deep learning of free
  boundary and stefan problems},\ }\href
  {https://doi.org/10.1016/j.jcp.2020.109914} {\bibfield  {journal} {\bibinfo
  {journal} {Journal of Computational Physics}\ }\textbf {\bibinfo {volume}
  {428}},\ \bibinfo {pages} {109914} (\bibinfo {year} {2021})}\BibitemShut
  {NoStop}%
\bibitem [{\citenamefont {Long}\ \emph {et~al.}(2026)\citenamefont {Long},
  \citenamefont {Zhou}, \citenamefont {Zhu}, \citenamefont {Dai},\ and\
  \citenamefont {Liu}}]{Long2026}%
  \BibitemOpen
  \bibfield  {author} {\bibinfo {author} {\bibfnamefont {Z.}~\bibnamefont
  {Long}}, \bibinfo {author} {\bibfnamefont {Q.}~\bibnamefont {Zhou}}, \bibinfo
  {author} {\bibfnamefont {A.}~\bibnamefont {Zhu}}, \bibinfo {author}
  {\bibfnamefont {D.}~\bibnamefont {Dai}},\ and\ \bibinfo {author}
  {\bibfnamefont {Y.}~\bibnamefont {Liu}},\ }\bibfield  {title} {\bibinfo
  {title} {Deep neural operator for free boundary problems},\ }\href
  {https://doi.org/10.1038/s42256-026-01233-9} {\bibfield  {journal} {\bibinfo
  {journal} {Nature Machine Intelligence}\ }\textbf {\bibinfo {volume} {8}},\
  \bibinfo {pages} {806} (\bibinfo {year} {2026})}\BibitemShut {NoStop}%
\bibitem [{\citenamefont {Laidler}(1984)}]{laidler1984development}%
  \BibitemOpen
  \bibfield  {author} {\bibinfo {author} {\bibfnamefont {K.~J.}\ \bibnamefont
  {Laidler}},\ }\bibfield  {title} {\bibinfo {title} {The development of the
  arrhenius equation},\ }\href@noop {} {\bibfield  {journal} {\bibinfo
  {journal} {Journal of chemical Education}\ }\textbf {\bibinfo {volume}
  {61}},\ \bibinfo {pages} {494} (\bibinfo {year} {1984})}\BibitemShut
  {NoStop}%
\bibitem [{\citenamefont {Varma}\ and\ \citenamefont
  {Simon}(2006)}]{Varma2006}%
  \BibitemOpen
  \bibfield  {author} {\bibinfo {author} {\bibfnamefont {S.}~\bibnamefont
  {Varma}}\ and\ \bibinfo {author} {\bibfnamefont {R.}~\bibnamefont {Simon}},\
  }\bibfield  {title} {\bibinfo {title} {Bias in error estimation when using
  cross-validation for model selection},\ }\href@noop {} {\bibfield  {journal}
  {\bibinfo  {journal} {BMC Bioinformatics}\ }\textbf {\bibinfo {volume} {7}},\
  \bibinfo {pages} {91} (\bibinfo {year} {2006})}\BibitemShut {NoStop}%
\bibitem [{\citenamefont {Cawley}\ and\ \citenamefont
  {Talbot}(2010)}]{Cawley2010}%
  \BibitemOpen
  \bibfield  {author} {\bibinfo {author} {\bibfnamefont {G.~C.}\ \bibnamefont
  {Cawley}}\ and\ \bibinfo {author} {\bibfnamefont {N.~L.~C.}\ \bibnamefont
  {Talbot}},\ }\bibfield  {title} {\bibinfo {title} {On over-fitting in model
  selection and subsequent selection bias in performance evaluation},\
  }\href@noop {} {\bibfield  {journal} {\bibinfo  {journal} {Journal of Machine
  Learning Research}\ }\textbf {\bibinfo {volume} {11}},\ \bibinfo {pages}
  {2079} (\bibinfo {year} {2010})}\BibitemShut {NoStop}%
\bibitem [{\citenamefont {Roberts}\ \emph {et~al.}(2017)\citenamefont
  {Roberts}, \citenamefont {Bahn}, \citenamefont {Ciuti}, \citenamefont
  {Boyce}, \citenamefont {Elith}, \citenamefont {Guillera-Arroita},
  \citenamefont {Hauenstein}, \citenamefont {Lahoz-Monfort}, \citenamefont
  {Schr{\"o}der}, \citenamefont {Thuiller}, \citenamefont {Warton},
  \citenamefont {Wintle}, \citenamefont {Hartig},\ and\ \citenamefont
  {Dormann}}]{Roberts2017}%
  \BibitemOpen
  \bibfield  {author} {\bibinfo {author} {\bibfnamefont {D.~R.}\ \bibnamefont
  {Roberts}}, \bibinfo {author} {\bibfnamefont {V.}~\bibnamefont {Bahn}},
  \bibinfo {author} {\bibfnamefont {S.}~\bibnamefont {Ciuti}}, \bibinfo
  {author} {\bibfnamefont {M.~S.}\ \bibnamefont {Boyce}}, \bibinfo {author}
  {\bibfnamefont {J.}~\bibnamefont {Elith}}, \bibinfo {author} {\bibfnamefont
  {G.}~\bibnamefont {Guillera-Arroita}}, \bibinfo {author} {\bibfnamefont
  {S.}~\bibnamefont {Hauenstein}}, \bibinfo {author} {\bibfnamefont {J.~J.}\
  \bibnamefont {Lahoz-Monfort}}, \bibinfo {author} {\bibfnamefont
  {B.}~\bibnamefont {Schr{\"o}der}}, \bibinfo {author} {\bibfnamefont
  {W.}~\bibnamefont {Thuiller}}, \bibinfo {author} {\bibfnamefont {D.~I.}\
  \bibnamefont {Warton}}, \bibinfo {author} {\bibfnamefont {B.~A.}\
  \bibnamefont {Wintle}}, \bibinfo {author} {\bibfnamefont {F.}~\bibnamefont
  {Hartig}},\ and\ \bibinfo {author} {\bibfnamefont {C.~F.}\ \bibnamefont
  {Dormann}},\ }\bibfield  {title} {\bibinfo {title} {Cross-validation
  strategies for data with temporal, spatial, hierarchical, or phylogenetic
  structure},\ }\href {https://doi.org/10.1111/ecog.02881} {\bibfield
  {journal} {\bibinfo  {journal} {Ecography}\ }\textbf {\bibinfo {volume}
  {40}},\ \bibinfo {pages} {913} (\bibinfo {year} {2017})}\BibitemShut
  {NoStop}%
\bibitem [{\citenamefont {Cunha}\ \emph {et~al.}(1998)\citenamefont {Cunha},
  \citenamefont {Oliveira},\ and\ \citenamefont {Oliveira}}]{Cunha1998}%
  \BibitemOpen
  \bibfield  {author} {\bibinfo {author} {\bibfnamefont {L.~M.}\ \bibnamefont
  {Cunha}}, \bibinfo {author} {\bibfnamefont {F.~A.~R.}\ \bibnamefont
  {Oliveira}},\ and\ \bibinfo {author} {\bibfnamefont {J.~C.}\ \bibnamefont
  {Oliveira}},\ }\bibfield  {title} {\bibinfo {title} {Optimal experimental
  design for estimating the kinetic parameters of processes described by the
  {W}eibull probability distribution function},\ }\href
  {https://doi.org/10.1016/S0260-8774(98)00085-5} {\bibfield  {journal}
  {\bibinfo  {journal} {Journal of Food Engineering}\ }\textbf {\bibinfo
  {volume} {37}},\ \bibinfo {pages} {175} (\bibinfo {year} {1998})}\BibitemShut
  {NoStop}%
\bibitem [{\citenamefont {Aliyari}\ \emph {et~al.}(2026)\citenamefont
  {Aliyari}, \citenamefont {{di Bari}}, \citenamefont {Meacci}, \citenamefont
  {Ratcliffe},\ and\ \citenamefont {Gray}}]{Aliyari2026}%
  \BibitemOpen
  \bibfield  {author} {\bibinfo {author} {\bibfnamefont {A.}~\bibnamefont
  {Aliyari}}, \bibinfo {author} {\bibfnamefont {V.}~\bibnamefont {{di Bari}}},
  \bibinfo {author} {\bibfnamefont {L.}~\bibnamefont {Meacci}}, \bibinfo
  {author} {\bibfnamefont {L.~P.~D.}\ \bibnamefont {Ratcliffe}},\ and\ \bibinfo
  {author} {\bibfnamefont {D.}~\bibnamefont {Gray}},\ }\bibfield  {title}
  {\bibinfo {title} {Semi-continuous process for rapid recovery of intact
  oleosomes via thermally accelerated swelling: Overcoming the activation
  energy barrier for seed imbibition},\ }\href
  {https://doi.org/10.1016/j.fbio.2026.108662} {\bibfield  {journal} {\bibinfo
  {journal} {Food Bioscience}\ }\textbf {\bibinfo {volume} {79}},\ \bibinfo
  {pages} {108662} (\bibinfo {year} {2026})}\BibitemShut {NoStop}%
\bibitem [{\citenamefont {Manz}\ \emph {et~al.}(2005)\citenamefont {Manz},
  \citenamefont {M{\"u}ller}, \citenamefont {Kucera}, \citenamefont {Volke},\
  and\ \citenamefont {Leubner-Metzger}}]{Manz2005}%
  \BibitemOpen
  \bibfield  {author} {\bibinfo {author} {\bibfnamefont {B.}~\bibnamefont
  {Manz}}, \bibinfo {author} {\bibfnamefont {K.}~\bibnamefont {M{\"u}ller}},
  \bibinfo {author} {\bibfnamefont {B.}~\bibnamefont {Kucera}}, \bibinfo
  {author} {\bibfnamefont {F.}~\bibnamefont {Volke}},\ and\ \bibinfo {author}
  {\bibfnamefont {G.}~\bibnamefont {Leubner-Metzger}},\ }\bibfield  {title}
  {\bibinfo {title} {Water uptake and distribution in germinating tobacco seeds
  investigated in vivo by nuclear magnetic resonance imaging},\ }\href
  {https://doi.org/10.1104/pp.105.061663} {\bibfield  {journal} {\bibinfo
  {journal} {Plant Physiology}\ }\textbf {\bibinfo {volume} {138}},\ \bibinfo
  {pages} {1538} (\bibinfo {year} {2005})}\BibitemShut {NoStop}%
\bibitem [{\citenamefont {Kikuchi}\ \emph {et~al.}(2006)\citenamefont
  {Kikuchi}, \citenamefont {Koizumi}, \citenamefont {Ishida},\ and\
  \citenamefont {Kano}}]{Kikuchi2006}%
  \BibitemOpen
  \bibfield  {author} {\bibinfo {author} {\bibfnamefont {K.}~\bibnamefont
  {Kikuchi}}, \bibinfo {author} {\bibfnamefont {M.}~\bibnamefont {Koizumi}},
  \bibinfo {author} {\bibfnamefont {N.}~\bibnamefont {Ishida}},\ and\ \bibinfo
  {author} {\bibfnamefont {H.}~\bibnamefont {Kano}},\ }\bibfield  {title}
  {\bibinfo {title} {Water uptake by dry beans observed by micro-magnetic
  resonance imaging},\ }\href {https://doi.org/10.1093/aob/mcl145} {\bibfield
  {journal} {\bibinfo  {journal} {Annals of Botany}\ }\textbf {\bibinfo
  {volume} {98}},\ \bibinfo {pages} {545} (\bibinfo {year} {2006})}\BibitemShut
  {NoStop}%
\bibitem [{\citenamefont {Glorot}\ and\ \citenamefont
  {Bengio}(2010)}]{Glorot2010}%
  \BibitemOpen
  \bibfield  {author} {\bibinfo {author} {\bibfnamefont {X.}~\bibnamefont
  {Glorot}}\ and\ \bibinfo {author} {\bibfnamefont {Y.}~\bibnamefont
  {Bengio}},\ }\bibfield  {title} {\bibinfo {title} {Understanding the
  difficulty of training deep feedforward neural networks},\ }in\ \href@noop {}
  {\emph {\bibinfo {booktitle} {Proceedings of the Thirteenth International
  Conference on Artificial Intelligence and Statistics}}},\ \bibinfo {series}
  {Proceedings of Machine Learning Research}, Vol.~\bibinfo {volume} {9}\
  (\bibinfo {year} {2010})\ pp.\ \bibinfo {pages} {249--256}\BibitemShut
  {NoStop}%
\bibitem [{\citenamefont {Kingma}\ and\ \citenamefont {Ba}(2015)}]{Kingma2015}%
  \BibitemOpen
  \bibfield  {author} {\bibinfo {author} {\bibfnamefont {D.~P.}\ \bibnamefont
  {Kingma}}\ and\ \bibinfo {author} {\bibfnamefont {J.}~\bibnamefont {Ba}},\
  }\bibfield  {title} {\bibinfo {title} {Adam: A method for stochastic
  optimization},\ }in\ \href@noop {} {\emph {\bibinfo {booktitle} {3rd
  International Conference on Learning Representations (ICLR)}}}\ (\bibinfo
  {year} {2015})\ \bibinfo {note} {arXiv:1412.6980}\BibitemShut {NoStop}%
\bibitem [{\citenamefont {Bradbury}\ \emph {et~al.}(2018)\citenamefont
  {Bradbury}, \citenamefont {Frostig}, \citenamefont {Hawkins}, \citenamefont
  {Johnson}, \citenamefont {Leary}, \citenamefont {Maclaurin}, \citenamefont
  {Necula}, \citenamefont {Paszke}, \citenamefont {VanderPlas}, \citenamefont
  {Wanderman-Milne},\ and\ \citenamefont {Zhang}}]{Bradbury2018}%
  \BibitemOpen
  \bibfield  {author} {\bibinfo {author} {\bibfnamefont {J.}~\bibnamefont
  {Bradbury}}, \bibinfo {author} {\bibfnamefont {R.}~\bibnamefont {Frostig}},
  \bibinfo {author} {\bibfnamefont {P.}~\bibnamefont {Hawkins}}, \bibinfo
  {author} {\bibfnamefont {M.~J.}\ \bibnamefont {Johnson}}, \bibinfo {author}
  {\bibfnamefont {C.}~\bibnamefont {Leary}}, \bibinfo {author} {\bibfnamefont
  {D.}~\bibnamefont {Maclaurin}}, \bibinfo {author} {\bibfnamefont
  {G.}~\bibnamefont {Necula}}, \bibinfo {author} {\bibfnamefont
  {A.}~\bibnamefont {Paszke}}, \bibinfo {author} {\bibfnamefont
  {J.}~\bibnamefont {VanderPlas}}, \bibinfo {author} {\bibfnamefont
  {S.}~\bibnamefont {Wanderman-Milne}},\ and\ \bibinfo {author} {\bibfnamefont
  {Q.}~\bibnamefont {Zhang}},\ }\href@noop {} {\bibinfo {title} {{JAX}:
  Composable transformations of {P}ython+{N}um{P}y programs}},\ \bibinfo
  {howpublished} {\url{http://github.com/google/jax}} (\bibinfo {year}
  {2018})\BibitemShut {NoStop}%
\bibitem [{\citenamefont {Efron}\ and\ \citenamefont
  {Tibshirani}(1993)}]{Efron1993}%
  \BibitemOpen
  \bibfield  {author} {\bibinfo {author} {\bibfnamefont {B.}~\bibnamefont
  {Efron}}\ and\ \bibinfo {author} {\bibfnamefont {R.~J.}\ \bibnamefont
  {Tibshirani}},\ }\href@noop {} {\emph {\bibinfo {title} {An Introduction to
  the Bootstrap}}}\ (\bibinfo  {publisher} {Chapman \& Hall},\ \bibinfo
  {address} {New York},\ \bibinfo {year} {1993})\BibitemShut {NoStop}%
\end{thebibliography}%
